\documentclass[11pt]{article}
\usepackage{a4wide}
\usepackage[PostScript=dvips]{diagrams}
\usepackage{latexsym}
\usepackage{amsmath,amssymb,stmaryrd}
\usepackage{pstricks,pst-node,pst-tree}
\usepackage{epsfig}
\usepackage{QED}
\usepackage{proof}
\usepackage[all]{xy}

\title{Information, Processes and Games}

\author{Samson Abramsky\\
Oxford University Computing Laboratory}
\date{}

\newcommand{\Prop}{\mathbb{P}}
\newcommand{\RV}{\mathsf{Rel}}
\newcommand{\DLbox}[1]{[ #1]}
\newcommand{\Act}{\mathsf{Act}}
\newcommand{\trans}[3]{#1 \stackrel{#2}{\lrarr} #3}
\newcommand{\otrans}[3]{#1 \stackrel{#2}{\Longrightarrow} #3}
\newcommand{\bang}{\mbox{!}}
\newcommand{\llpar}{\bindnasrepma}
\newcommand{\llto}{\multimap}
\newcommand{\llnot}[1]{#1^{\bot}}
\newcommand{\pfun}{\rightharpoonup}
\newcommand{\CL}{\mathbf{CL}}
\newcommand{\babs}[2]{\lambda^{\ast} #1 . \, #2}
\newcommand{\CC}{\mathbf{C}}
\newcommand{\DD}{\mathbf{D}}

\newcommand{\SSS}{\mathbf{S}}
\newcommand{\BB}{\mathbf{B}}
\newcommand{\BBB}{\mathbf{B}}
\newcommand{\KK}{\mathbf{K}}
\newcommand{\II}{\mathbf{I}}
\newcommand{\FF}{\mathbf{F}}
\newcommand{\WW}{\mathbf{W}}

\newcommand{\CCC}{\mathbf{C}}

\newcommand{\ap}{\cdot}

\newcommand{\linimpl}{\multimap}
\newcommand{\llbang}{\, !}
\newcommand{\Pos}{\mathsf{Pos}}
\newcommand{\pair}{\mathsf{p}}
\newcommand{\Split}{\mathsf{s}}
\newcommand{\spar}{\, | \,}
\newtheorem{theorem}{Theorem}[section]

\newtheorem{definition}[theorem]{Definition}

\newcommand{\fact}{\mathsf{fact}}
\newcommand{\cond}[3]{\mathbf{if} \; #1 \; \mathbf{then} \; #2 \; \mathbf{else} \; #3}
\newcommand{\while}[2]{\mathbf{while} \; #1 \; \mathbf{do} \; #2}

\newcommand{\Nat}{\mathbb{N}}
\newcommand{\Bool}{\mathbb{B}}
\newcommand{\Term}{\mathbf{1}}
\newcommand{\Natbot}{\mathbb{N}_{\bot}}
\newcommand{\Boolbot}{\mathbb{B}_{\bot}}
\newcommand{\Termbot}{\mathbf{1}_{\bot}}

\newcommand{\Sierp}{\mathbb{O}}
\newcommand{\true}{\mathsf{tt}}
\newcommand{\false}{\mathsf{ff}}

\newcommand{\eqdef}{\equiv}
\newcommand{\sqle}{\sqsubseteq}

\newcommand{\rarr}{\rightarrow}
\newcommand{\lrarr}{\longrightarrow}
\newcommand{\ochain}{$\omega$-chain~}
\newcommand{\olub}{\bigsqcup_{n \in \omega}}

\newcommand{\ie}{\textit{i.e.}~}
\newcommand{\Pow}[1]{\mathcal{P}(#1)}

\newcommand{\pfn}{\rightharpoonup}
\newcommand{\Pfn}[2]{\mathbf{Pfn}(#1, #2)}

\newcommand{\Siginf}{\Sigma^{\infty}}
\newcommand{\lfp}[1]{\mathsf{lfp}(#1)}

\newcommand{\funsp}[2]{[#1 \longrightarrow #2]}

\newcommand{\varempty}{\varnothing}

\newcommand{\IDom}{\mathbb{I}[0, 1]}

\newarrow{incarrow}C--->

\begin{document}
\maketitle

\begin{abstract}
We survey the prospects for an Information Dynamics which can serve as the basis for a fundamental theory of information, incorporating qualitative and structural as well as quantitative aspects. We motivate our discussion with some basic conceptual puzzles: how can information increase in computation, and what is it that we are actually computing in general? Then we survey a number of the theories which have been developed within Computer Science, as partial exemplifications of the kind of fundamental theory which we seek: including Domain Theory, Dynamic Logic, and Process Algebra. We look at recent work showing new ways of combining quantitative and qualitative theories of information, as embodied respectively by Domain Theory and Shannon Information Theory. Then we look at Game Semantics and Geometry of Interaction, as examples of dynamic models of logic and computation in which  information flow and interaction are made central and explicit. We conclude by looking briefly at some  key issues for future progress.
\end{abstract}

\tableofcontents

\section{Prelude: Some Basic Puzzles}

Before attempting a conventional introduction to this article, we shall formulate some basic puzzles which may serve as motivation for, and an indication of, some of the themes we shall  address.

\subsection{Does Information Increase in Computation?}
Let us begin with a simple-minded question:
\begin{center}
\fbox{\textbf{Why do we compute?}}
\end{center}
The natural answer is: \emph{to gain information} (which we did not previously have)!
But how is this possible?\footnote{Indeed, I was once challenged on this point by an eminent physicist (now knighted), who demanded to know how I could speak of information increasing in computation when Shannon Information theory tells us that it cannot! My failure to answer this point very convincingly at the time led me to continue to ponder the issue, and eventually gave rise to this discussion.}
\begin{description}
\item[Problem 1:] Isn't the output \emph{implied} by the input?
\item[Problem 2:] Doesn't this contradict the second law of thermodynamics?
\end{description}

\paragraph{A logical form of Problem 1}
This problem  lies adjacent to another one at the roots of logic. If we extract logical consequences of axioms, then surely the answer was already there implicitly in the axioms; what has been added by the derivation? Since computation can itself, via the Curry-Howard isomorphism \cite{Cur58,How80,GLT}, be modelled as performing \emph{Cut elimination} on proofs, or \emph{normalization} of terms, the same question can be asked of computation. A normal form which is presented as the result of a computation is logically \emph{equal} to the term we started with: 
\[ M \longrightarrow^* N \quad \Longrightarrow \quad \llbracket M \rrbracket = \llbracket N \rrbracket . \]
so what has been added by computing it? 

The same issue can be formulated in terms of the logic programming paradigm, or of querying a relational database \cite{CGT90}: in both cases, the result of the query is a logical consequence of the data- or knowledge-base.

The challenge here is to build a useful theory which provides convincing and helpful answers to these questions. We simply make some preliminary observations.
Note that normal forms are in general \emph{unmanagably big} \cite{Vor97}.
Useful output has two aspects:
\begin{itemize}
\item Making information explicit---\ie extracting the normal form.
\item Data reduction---getting rid of a lot of the information in the input.
\end{itemize}
(Note that it is \emph{deletion of data} which creates thermodynamic cost in computation \cite{Lan00}).
Thus we can say that much (or all?)  of the actual usefulness of computation lies in getting rid of the hay-stack, leaving only the needle.

\paragraph{Problem 2: Discussion}
While information is presumably conserved in the \emph{total} system, there can be information flow between, and information increase in, \emph{subsystems}.
(A body can gain heat from its environment). More precisely, while the entropy of an isolated (total) system cannot decrease, a sub-system \emph{can} decrease its entropy by consuming energy from its environment.

Thus if we wish to speak of information flow and increase, this must be done relative to subsystems. Indeed, the fundamental objects of study should be \emph{open systems}, whose behaviour must be understood in relation to an external environment. Subsystems which can \emph{observe} incoming information from their environment, and \emph{act} to send information to their environment, have the capabilities of \emph{agents}.

\paragraph{Observer-dependence of information increase}
Yorick Wilks (personal communication) has suggested the following additional twist. Consider an equation such as
\[ 3 \times 5 = 15 . \]
The forward direction $3 \times 5 \rightarrow 15$ is obviously a natural direction of computation, where we perform a multiplication. But the reverse direction $15 \rightarrow 3 \times 5$ is also of interest --- finding the prime factors of a number! So the \emph{direction of possible information increase} must  be understood as relative to the observer or user of the computation.\footnote{Formally, this can be understood in terms of different choices of normal forms. For a general perspective on rewriting as a computational paradigm, see \cite{BN,Ter}.}

\paragraph{Moral:} Agents and their interactions are intrinsic to the study of information flow and increase in computation.
The classical theories of information \emph{do not  reflect this adequately}.

\subsection{What Function Does the Internet Compute?}

Our second puzzle reflects the changing conception of computation which has been developing within Computer Science over the past three decades. The traditional conception of computation is that we compute an output as a function of an input, by an algorithmic process. This is the basic setting for the entire field of algorithms and complexity, for example. So \emph{what} we are computing is clear --- it is a function.\footnote{We may, if we are willing to countenance non-deterministic or probabilistic computation, be willing to stretch this functional paradigm to accomodate  relations or stochastic relations of some kind. These are minor variations, compared to the shift to a fully-fledged dynamical perspective.}  But the reality of modern computing: distributed, global, mobile, interactive, multi-media, embedded, autonomous, virtual, pervasive, \ldots\footnote{See e.g.~\cite{Mil06a,Mil06b}.}  --- forces us to confront the limitations of this viewpoint.

Traditionally, the \emph{dynamics} of computing systems --- their unfolding behaviour in space and time --- has been a mere means to the end of computing the function which specifies the algorithmic problem which the system is solving.\footnote{Insofar as the dynamics has been of interest, it has been in quantitative terms, counting the resources which the algorithmic process consumes --- leading of course to the notions of algorithmic complexity. Is it too fanciful to speculate that the lack of an adequate structural theory of processes may have been an impediment to fundamental progress in complexity theory?} In much of contemporary computing, the situation is reversed: the \emph{purpose} of the computing system is to exhibit certain behaviour. The \emph{implementation} of this required behaviour will seek to reduce various aspects of the specification to the solution of standard algorithmic problems.

\begin{center}
\fbox{\textbf{What does the Internet compute?}}
\end{center}
Surely not a mathematical function \ldots

\paragraph{Moral:} We need a theory of the dynamics of informatic processes, of \emph{interaction}, and \emph{information flow}, as a basis for answering such fundamental questions as :
\begin{itemize}
\item \emph{What} is computed? 
\item What \emph{is} a process? 
\item What are the analogues to
Turing-completeness and universality when we are concerned with processes and their behaviours, rather than the functions which they compute?
\end{itemize}

\section{Introduction: Matter and Method}

Philosophers of science are concerned with explaining various aspects of science, and often, moreover, with viewing science as a kind of gold-mine of philosophical opportunity. The direction in both cases is \emph{philosophy from science}. For a theoretical or mathematical scientist, the primary inclination is often to see conceptual analysis as a preliminary to a more technical investigation, which may lead to a new theoretical development. In short: \emph{science from philosophy}. This article is written mainly in the latter spirit, from the stand-point of Theoretical Computer Science, or perhaps more broadly ``Theoretical Informatics'': a --- still largely putative --- general science of information. That being said, we hope that our conceptual discussions may also provide some useful grist to the philosopher's mill.

\subsection{Towards Information Dynamics}
The best-known existing mathematical theories of information are (largely) \emph{static} in nature. That is, they do not explicitly describe informatic processes and information flow, but rather certain \emph{invariants} of these processes and flows. There is by now ample experience from Computer Science which indicates that it is fruitful, and eventually necessary, to develop fully-fledged dynamical theories. 
We shall try to map some steps in this direction. 

We begin by reviewing some of the theories developed in Computer Science which form the background for our discussion.
Then we consider  another important issue in theories of information: the distinction between \emph{qualitative} and \emph{quantitative} theories, and how they can be reconciled --- or, more positively, combined. Our discussion here will still be at the level of \emph{static} theories. We then go on to consider dynamic theories proper. 

This article is well outside the author's usual remit as a researcher. While it is clearly not a contribution to philosophy, it cannot be said to be the usual kind of conceptually-oriented overview of a scientific field which one might find in such a Handbook (and of which there are some fine examples in the present volume) either; not least for the reason that the scientific field we are attempting to overview does not exist yet, in a fully realized form at any rate. Rather, the main purpose of this article is to play some small part in helping this field to come into being.

What, then, is this nascent field? We would like to use the term \emph{Information Dynamics}, which was proposed some time ago by Robin Milner, to suggest how the area of Theoretical Computer Science usually known as ``Semantics'' might emancipate itself from its traditional focus on interpreting the syntax of pre-existing programming languages, and become a more autonomous study of the fundamental structures of Informatics.\footnote{Robin Milner has also written several articles in the same general spirit as this one, notably \cite{Mil96}.}
The development of such a field would transform our scientific vision of Information, and give us a whole new set of tools for thinking about it. Hence its relevance for any attempt to develop a Philosophy of Information.

Rather than a developed field of Information Dynamics, with some consensus as to what its fundamental notions and methods are, what we have at present are some \emph{partial exemplifications}; some theories which have been shown to work well over certain ranges of applications, and which exhibit both conceptual and mathematical depth. Our approach to conveying the current state of the art, and indicating  the major objectives visible from where we stand now, is necessarily largely based on describing (some of) these current theories. 
The obvious danger with this approach is that this article will appear to be a disjointed series of descriptions of various formalisms. We have probably not succeeded in avoiding this completely---despite the author's best efforts. But we regard the expository aspect of this article as important in itself. The theories we shall expound deserve to be known in wider circles than they presently are. And our discussions of Domain Theory, Game semantics and Geometry of Interaction delve more into conceptual issues, while minimizing the level of technical detail, than other accounts of which we are aware.

\subsection{Some Themes}
To assist the reader in keeping their bearings, we mention some of the main themes which will thread through our discussion:
\begin{description}
\item[Information Increase in Computation] We compute in order to gain information: but how is this possible, logically or thermodynamically? How can it be reconciled with the point of view of Information Theory? How does information increase appear in the various extant theories? This will be an important explicit theme in our discussion of background theories in Section~3, and particularly in Section~4. Obtaining a good account in the context of dynamic theories, as exemplified by those presented in Sections~5 and~6, is a key desideratum for future work.

\item[ Unifying Quantitative and Qualitative Theories of Information] 
We mainly discuss this explicitly in Section~4, where we describe some striking recent progress which has been achieved by Keye Martin and Bob Coecke, in the setting of current static theories of information (Scott Domain Theory and Shannon Information Theory). A similar development in the setting of the dynamic theories described in Sections~5 and~6 is a major objective for future research.

\item[Information Dynamics: Logic and Geometry]
We introduce Game Semantics and Geometry of Interaction in Sections~5 and~6 as substantial partial exemplications of Information Dynamics. They have strong connections to both Logic and Geometry, and form a promising new bridge between these two fields. While we shall not be able to do full justice to these topics, we hope at least to raise the reader's awareness of these developments, and to provide pointers into the literature.

\item[The Power of Copying, and Logical Emergence]
This is mainly developed in Section~6, in the context of Geometry of Interaction-type  models. The theme here is to look at how logically complex behaviour can emerge from very simple ``copy-cat processes'', showing the power of interaction. The links between the interactive and geometric points of view become very clear at this basic level.

\end{description}
One theme which we have, regretfully, omitted is that of  the emerging connections with Physics, in particular with \textbf{Quantum Information and Computation}. Here there is already much to say (see e.g. \cite{Abramsky2002,AbrCoe1,AbrCoe2}). We have not included this material simply for lack of the appropriate physical resources of space, time and energy.

\paragraph{Acknowledgements}
A number of people kindly agreed to read a draft of this article, and provided very perceptive and helpful comments: Adam Brandenburger, Jeremy Butterfield, Robin Milner and Yorick Wilks. I would like to express my warmest appreciation for their input. My thanks also to Jan van Eijck, who commented on the paper on behalf of the Editors, and produced two rounds of useful comments. Special thanks to Johan van Benthem, who asked me to write the article in the first place, and whose encouragement, suggestions and gentle reminders have kept me on track.

\section{Some Background Theories}

Following our previous discussion, we can classify theories of information along two axes: as static or dynamic, and as qualitative or quantitative. We list some examples in the following table.
\begin{center}
\begin{tabular}{l|ll}
& Static & Dynamic \\ \hline
Qualitative & Domain Theory, & Process Algebra \\
&  Dynamic Logic & \\
Quantitative & Shannon Information theory & \\
\end{tabular}
\end{center}
It may seem strange to list Dynamic Logic as a static theory --- and indeed, not everyone would agree with this classification! We regard it as static because it considers input-output relations only, and not the structure of the processes which realize these relations. The distinction we have in mind will become clearer when we go on to discuss Process Algebra.

Shannon Information theory is discussed in detail in another Chapter of this Handbook. In this Section, we shall give brief overviews of the other three theories listed above, which have all been developed within Computer Science---Domain Theory and Dynamic Logic originating in the 1970's, and Process Algebra in the  1980's.

It may be useful to give a timeline for some of the seminal publications:
\begin{center}
\begin{tabular}{l|lll}
1948 & Claude Shannon & \textit{A Mathematical Theory of Communication} & Information Theory \\  \hline
1963 &  Saul Kripke & \textit{Semantical Considerations on Modal Logic} & Kripke Structures \\  \hline
1969 &  Dana Scott & \textit{Outline of a Mathematical Theory of Computation} & Domain Theory \\  
& Tony Hoare & \textit{An Axiomatic Basis for Computer Programming} & Hoare Logic \\ \hline
1976 & Vaughan Pratt & \textit{Semantical Considerations on Floyd-Hoare Logic} & Dynamic Logic \\  
& Johan van Benthem & \textit{Modal Correspondence Theory} & Bisimulation \\ \hline
1980 & Robin Milner & \textit{A Calculus of Communicating Systems} & Process Algebra \\  \hline
\end{tabular}
\end{center}
The work on Game Semantics and Geometry of Interaction to be covered in Sections~5 and 6 comes from the 1990's. As always, a full intellectual history is complex, and we shall not attempt this here.

We shall devote rather more space to Domain Theory than to the other two theories, for the following reasons:
\begin{itemize}
\item Domain Theory is more intrinsically and explicitly a theory of information than Dynamic Logic or Process Algebra, and will figure significantly in our subsequent discussions.
\item The other theories will receive some coverage elsewhere in this Handbook, notably in the Chapter by Baltag and Moss.
\end{itemize}

\subsection{Domain Theory}
Domain Theory was introduced by Dana Scott \textit{c.}~1970 \cite{scott:outline} as a mathematical foundation for the denotational semantics of programming languages which had been pioneered by Christopher Strachey. A \emph{domain} is a partially ordered structure $(D, \sqle)$. The best intuitive reading of elements of $D$ is as \emph{information states}.
We pass immediately to some illustrative examples.

\subsubsection{Examples of Domains}

\subparagraph{Flat Domains} Given a set $X$, we can form a domain $X_{\bot}$ by adjoining an element $\bot \not\in X$, and defining an order by
\[ x \sqle y \;\; \Longleftrightarrow \;\; x = \bot \; \vee \; x = y . \]
Frequently used examples : $\Natbot$, $\Boolbot$, $\Sierp = \Termbot$.
Here $\Nat = \{ 0, 1, 2, \ldots \}$, the set of \emph{natural numbers}; $\Bool = \{ \true , \false \}$, the set of \emph{booleans}; and $\Term = \{ \ast \}$, an (arbitrary) one-element set.

We can use such flat domains to model computations in terms of very simple \emph{processes of information increase}. Thus a (possibly non-terminating) natural number computation can be modelled in $\Natbot$ in the following sense. Initially, no output has been produced. This ``zero information state'' is represented by the bottom element $\bot$. If the computation terminates, a natural number $n$ is produced. Thus we obtain the ``process''
\[ \bot \sqle n . \]
The case where no output is ever produced is captured by the ``stationary process'' $\bot$, which we can view more ``dynamically'' as
\[ \bot \sqle \bot \sqle \cdots \]

\subparagraph{Streams}
Now consider the scenario where we have an unbounded or potentially infinite tape (much as for the \emph{output tape} of a Turing machine), on successive squares of which symbols from some finite alphabet $\Sigma$ can be printed. This computational scenario is naturally modelled by the domain $\Siginf$,  the set of finite and infinite sequences of elements of $\Sigma$. This is ordered by \emph{prefix}: $x \sqle y$ if $x = y$, or $x$ is finite, and for some (finite or infinite) sequence $z$, $xz = y$.
Example:
\[ \langle 0 \rangle \sqle  \langle 0, 0 \rangle \sqle  \langle 0, 0, 0 \rangle \sqle \cdots \sqle 0^{\omega} \]
where $0^{\omega}$ is the infinite sequence of $0$'s.

This example shows the ability of domain theory to model infinite computations as \emph{limits} of processes of information increase, where at each stage in the process the information state is finite.

It is important to distinguish a finite stream in this domain from a finite list as a standard programming data structure, e.g.~in LISP. A finite list in standard usage is a \emph{complete}, informationally perfect object, just like a natural number in our previous example. A finite stream, by contrast, has a ``sting in the tail''; a potentially infinite computation to determine what the remaining elements to be printed on the output tape will be. Thus a finite stream in the above domain is an informationally incomplete object, which can be extended to a more defined stream, which it then approximates.

\subparagraph{The Interval Domain}
Now suppose our computational scenario is that we are computing a real number in the unit interval $[0, 1]$. Clearly we can only compute to finite precision in finite time (and with finite resources), so we are \emph{forced} to consider a scenario of approximation. The appropriate domain here is $\IDom$, consisting of all closed non-empty intervals $[a, b]$ where $0 \leq a \leq b \leq 1$. We read an interval $[a, b]$ as expressing our current state of information about the real $r \in [0, 1]$ we are computing, namely that $a \leq r \leq b$. The ordering is by reverse inclusion of intervals, or equivalently by
\[ [a, b] \sqle [c, d] \;\; \Longleftrightarrow \;\; a \leq c \; \wedge \; d \leq b . \]
This corresponds to \emph{refinement} of our information state to a more accurate determination of the location of the ideal element $r$. Note that the case $[r, r]$ is allowed, for any $r \in [0, 1]$. In fact, this embeds the unit interval into the interval domain as the set of \emph{maximal elements} of $\IDom$.
Note that for any real number $r \in [0, 1]$, there is a process of information increase
\[ [0, 1] \sqle [a_{1}, b_{1}] \sqle [a_{2}, b_{2}] \sqle \cdots \]
where $a_{n+1} = a_{n}$ and $b_{n+1} = (a_{n} + b_{n})/2$ if $r$ is in the left half-interval of $[a_{n}, b_{n}]$, and $a_{n+1} = (a_{n} + b_{n})/2$ and $b_{n+1} = b_{n}$ if $r$ is in the right half-interval. Clearly $r$ is the supremum of the $a_{n}$ and the infimum of the $b_{n}$. Thus every real can be computed as the limit of a process of information increase where at each finite stage of the process the interval has rational end-points, and hence is a finitely representable information state.\footnote{We are glossing over some technical subtleties here. The interval domain is a basic example of a \emph{continuous domain}---the only one we shall encounter in this brief sketch of domain theory. This means that ``finiteness'' does not have the same ``absolute'' status in this case that it does in our other examples.(Formally, intervals with rational end-points are not \emph{compact}.) Nevertheless, these finitely representable intervals do play a natural role in the \emph{effective presentation} of the domain, and the example is an important one for conveying the basic intuitions of Domain Theory. See \cite{AJ,compendium} for extensive coverage of continuous domains.}

\subparagraph{Partial Functions} A somewhat more abstract example is provided by
the set $\Pfn{X}{Y}$ of partial functions from $X$ to $Y$, ordered by inclusion. To see how this can be used in computational modelling, consider the recursive  definition
of the factorial function:
\[ \fact (n) = n! = n \cdot (n-1) \cdot (n-2) \cdots 2 \cdot 1 . \]
\[ \fact (n) = \cond{n=0}{1} {n \times \fact (n-1)} . \]
We can understand this recursive definition as specifying a process of information increase over the domain $\Pfn{\Nat}{\Nat}$. Initially, we are at the zero information state (least element of the domain) $\varnothing$; we know nothing about which ordered pairs are in the graph of the function being defined recursively. Inspection of the base case of the recursion (where $n=0$) allows us to deduce that the pair $(0, 1)$ is in the graph of the function. Once we know this, we can infer that in the case $n=1$,
\[ \fact (1) = 1 \times \fact (0) = 1 \times 1 = 1. \]
Thus the process of information increase proceeds as follows:
\[ \varempty \subseteq \{ (0, 1) \} \subseteq \{ (0, 1), (1, 1) \} \subseteq \cdots \]
We can see inductively that the $n$'th term in this sequence will give the values of factorial on the arguments from $0$ to $n-1$; and the \emph{least upper bound} of this increasing sequence, given simply by its union, will be the factorial function.

\subsubsection{Technical Issues}
These examples serve to motivate a number of additional \emph{axioms for domains}. There is in fact no unique axiom system for domains. We shall mention the most fundamental forms of such axioms.
\paragraph{Completeness}
As we have seen, an essential point of Domain Theory is to allow the description of infinite computations or computational objects as \emph{limits} of processes of information increase. A corresponding property of completeness of domains is required, to ensure that a well-defined unique limit exists for every such process. Such limits are expressed as \emph{least upper bounds} in order-theoretic terms. The idea is that for a process
\[ d_{0} \sqle d_{1} \sqle d_{2} \sqle \cdots \]
the limit should contain \emph{all} the information produced at any stage of the process; and \emph{only} the information produced by some stage of the process. The first point implies that the limit should be an upper bound; the second, that it should be the \emph{least} upper bound.

Which class of increasing sets should be regarded as processes of information increase? The most basic class, which has figured in all our examples to date, is that of \emph{increasing sequences}, or ``$\omega$-chains'' in the usual technical parlance. The axiom requiring completeness for all such chains, which picks out the class of ``$\omega$-complete partial orders'', or $\omega$-cpos for short, is often used in Domain Theory. We shall henceforth assume  that all domains we consider are $\omega$-cpos. Sometimes completeness for a larger class of sets, the \emph{directed sets}, is used. This reflects technical issues akin to the distinction in Topology between sequential completeness and completeness for nets or ultrafilters, and we shall not pursue this here.

\paragraph{Least Elements}
All our examples have had a least element: $\bot$ for flat domains, the empty stream for $\Siginf$, the unit interval $[0, 1]$ for $\IDom$, and the empty set for $\Pfn{X}{Y}$. This provides a zero information point, and hence a canonical starting point for processes of information increase. Mathematically, least elements are essential for the least fixed point theorem which we shall  encounter shortly. There are schemes for Domain Theory in which domains (or ``pre-domains'') are not required to have least elements in general, but they always  enter the theory at crucial points, sometimes through a general operation of adjoining a least element to a predomain to form a domain (``lifting'').

\paragraph{Approximation}
The intuition developed through our examples for how general elements of the domain can be approximated by others, which may in particular be of finite character, is captured formally by requiring domains to be \emph{algebraic} or \emph{continuous}. We shall not develop these notions here, but will simply note for our examples:
\begin{itemize}
\item For flat domains such as $\Natbot$, we can regard \emph{all} elements as of finite character.
\item Every stream in $\Siginf$ can be realized as the least upper bound of an increasing sequence of \emph{finite} streams.
\item Every real in $[0, 1]$, and more generally every interval in $\IDom$,  can be realized as the least upper bound of an increasing sequence of intervals with \emph{rational} end-points.
\item Every partial function in $\Pfn{X}{Y}$, and in particular every total function from $X$ to $Y$, where $X$ and $Y$ are countable, can be realized as the least upper bound of an increasing sequence of \emph{finite} partial functions. (The case where $X$ is uncountable is a typical example where we would naturally resort to general directed sets rather than sequences.)
\end{itemize}

\subsubsection{Conceptual Issues}
\paragraph{Why Partial Orders?}
Having developed some examples and intuitions, we now re-examine the basic concept of domains as partial orders $(D, \sqle)$. If we think of the elements of $D$ as information states, the way we articulate this structure is \emph{qualitative} in character. That is, we don't ask \emph{how much} information a given state contains, but rather a relational question: does one state convey \emph{more} information than another? We read $d \sqle e$ as ``$e$ conveys at least as much information as $d$''. If we consider the partial order axioms with this reading:
\begin{center}
\begin{tabular}{ll}
\textbf{Reflexivity}  & $x \sqsubseteq x$ \\
\textbf{Transitivity} &
$x \sqle y \;\; \wedge \;\; y \sqle z \quad \Longrightarrow \quad x \sqle z$ \\
\textbf{Anti-Symmetry} & $x \sqle y \;\; \wedge \;\; y \sqle x \quad \Longrightarrow \quad x = y$.
\end{tabular}
\end{center}
then Reflexivity is clear; and Transitivity also very natural. Anti-Symmetry can be seen as embodying an important \emph{Principle of Extensionality}: if two states convey the \emph{same} information, they are regarded as \emph{equal}.

\paragraph{States of What?}
We have been using the term ``information state'' to convey the intuition for what the elements of a domain represent. In fact, there is a certain creative ambiguity lurking here, between two interpretations of what these are states \emph{of}:
\begin{itemize}
\item We may think of states of a \emph{computational system} in itself, characterized in terms of the information they contain, as an ``intrinsic'' or ``objective'' property of the system, independently of any observer.
\item We may implicitly introduce an \emph{observer} of the system, and understand the information content of a system in terms of  the observer's state of information about it.
\end{itemize}
In the first reading, we  think of the partial elements of the domain in an ontological way, as necessary extensions to our universe of discourse to represent the range of possible outputs of computational systems which may run for ever, and may fail to terminate or to produce information beyond some finite stage of the computation. In the second reading, we are thinking  epistemologically: what information can the observer gain about the computation.

In fact, both readings are useful---and are widely used. It is very common to slip without explicit mention from one to the other---nor, for the technical purposes of the theory, does this seem to do any harm. Mathematically, this distinction can be related to the duality between \emph{points} and \emph{properties}, in the sense of Stone-type dualities: the duality between the points of a topological space, and its basic ``observable properties''---the open sets \cite{Joh82}. The particular feature of domains which allows this creative ambiguity between points and properties to be used so freely without incurring any significant conceptual confusions or overheads is that \emph{basic points and basic properties (or observations) are essentially the same things}. We explain this in terms of an example. Consider a finite steam $s$ in $\Siginf$. On the one hand, this can be viewed as a point, \ie as an element of the domain --- which may be produced by some system which computes the elements of $s$ in  finite time, and then continues to run forever without producing any more output. On the other hand, we may view this finite stream $s$ as a property: the property satisfied by any system with output stream $t$ such that $s \sqle t$. It is a \emph{finitely observable property}, since we can tell whether a system satisfies it after only a finite time spent observing the system. Whether we take $\Siginf$ as the space of points $X$ generated as limits of increasing sequences of finite streams, or as the ``logic'' (or open-set lattice) $L$ of properties generated by the basic observations given by finite streams, we get the same thing: the topology of $X$ will be $L$, and the space of points generated (as completely prime filters) over $L$ will be $X$. This is Stone duality. An extensive development of Stone duality for Domain Theory has been given in \cite{Abramsky91}; see also \cite{AJ,Zha91,BK99}.

In fact, we would argue that it is hard to avoid the epistemic stance entirely. For example, the plausibility of something as basic as the Anti-Symmetry axiom is much greater if we think in terms of an observer. Much of the conceptual power of Domain Theory comes from the idea that it articulates how we can approximate infinite ideal objects by processes which use only finite resources at each finite stage.

\paragraph{Static or Dynamic?}
Another subtle underlying issue which is not usually made explicit is that Domain Theory is \emph{a static theory resting on dynamic intuitions}. Indeed, we have motivated the theory in terms of certain \emph{processes of information increase}. Processes happen in time; thus time is present implicitly in Domain Theory. This underlying temporality can be developed more explicitly within the Domain Theoretic framework:
\begin{itemize}
\item One can add axioms to the basic ones for domains to pick out those domains which are \emph{concrete} \cite{KP78}, in the sense that we can understand information increase in terms of a temporal flow of events. Now the ordering is not simply one of information content, but involves an idea of \emph{causality}, so that some events \emph{must} temporally precede others. This leads to notions of \emph{event structures} \cite{NPW81}, which have been applied to the study of concurrent processes. Very similar structures have shown up recently in Theoretical Physics, in the Causal Sets approach to quantum gravity \cite{Sor}.
\item In some remarkable recent work, Domain Theoretic tools are used to characterize globally hyperbolic space-time manifolds in terms of their causal orderings \cite{MP06}.
\end{itemize}
However, it should be said that most of the applications of Domain Theory in denotational semantics are carried out at a much higher level of abstraction, where temporality appears only in the most residual form. This arises from the fact that computations or programs are modelled in the Domain-Theoretic denotational framework essentially as \emph{functions} from inputs to outputs.

\subsubsection{Continuous Functions}
We now consider the appropriate notion of function between domains. Let $D$, $E$ be $\omega$-complete partial orders. A function $f : D \rightarrow E$ is \emph{monotonic} if, for all $x, y \in D$:
\[ x \sqle y \;\; \Longrightarrow \;\; f(x) \sqle f(y) . \]
It is \emph{continuous} if it is monotonic, and for all $\omega$-chains $(x_{n})_{n \in \omega}$ in $D$:
\[ f (\bigsqcup_{n \in \omega} x_{n}) = \bigsqcup_{n \in \omega} f(x_{n}) . \]

\paragraph{Examples}
We consider a number of examples of functions $f : \Siginf \rightarrow \Boolbot$, where $\Sigma = \{ 0, 1\}$.
\begin{enumerate}
\item $f(x) = \true$ if $x$ contains a 1, $f(x) = \bot$ otherwise.
\item  $f(x) = \true$ if $x$ contains a 1, $f(0^{\infty}) = \false$, $f(x) = \bot$ otherwise.
\item  $f(x) = \true$ if $x$ contains a 1, $f(x) = \false$ otherwise.
\end{enumerate}
Of these: (1) is continuous, (2) is monotonic but not continuous, and (3) is not monotonic.

As these examples indicate, the conceptual basis for monotonicity is that \emph{the information in Domain Theory is positive; negative information is not regarded as stable observable information}. That is, if we are at some information state $s$, then for all we know, $s$ may still increase to $t$, where $s \sqle t$. This means that if we decide to produce information $f(s)$ at $s$, then we must produce all this information, and possibly more, at $t$, yielding $f(s) \sqle f(t)$. Thus we can only make decisions at a given  information state which are stable under every possible information increase from that state. This idea is very much akin to the use of partial orders in Kripke semantics for Intuitionistic Logic, in particular in connection with the interpretation of negation in that semantics.
The continuity condition, on the other hand, reflects the fact that a computational process will only have access to a finite amount of information at each finite stage of the computation. If we are provided with an infinite input, then any information we produce as output at any finite stage can only depend on some finite observation we have made of the input. This is reflected in one of the inequations corresponding to continuity:
\[ f (\bigsqcup_{n \in \omega} x_{n}) \sqle \bigsqcup_{n \in \omega} f(x_{n})  \]
which says that the information produced at the limit of an infinite process of information increase is no more than what can be obtained as the limit of the information produced at  the finite stages of the process.
Note that the ``other half'' of continuity
\[ \bigsqcup_{n \in \omega} f(x_{n}) \sqle f (\bigsqcup_{n \in \omega} x_{n}) \]
follows from monotonicity.

Note by the way how this discussion is permeated with the epistemic stance. Continuous functions produce \emph{points} as outputs on the basis of \emph{observations} they make of their inputs.  Thus the duality between these two points of view plays a basic r\^ole in our very \emph{understanding} of continuous functions.\footnote{Mathematically, this duality appears in the guise of the compact-open topology for function spaces. We can think of open sets in functions spaces as observations which can be made on functions viewed as black boxes. Dually to the point of view of the function, which \emph{observes} an input and \emph{produces} an output, a \emph{function environment} must produce an input (a point, or in more general topological situations, a compact set), and observe the corresponding output.} This  can be (and often is) glossed over in Domain Theory, by virtue of the coincidence of finite points and finite properties which we have already discussed.

\subsubsection{The Fixpoint Theorem}
We now consider a simple but powerful and very widely applicable theorem, which is one of the main pillars of Domain Theory, since by virtue of this result it provides a general setting in which  recursive definitions can be understood.\footnote{A \emph{fixed point} of a function $f : X \lrarr X$ is an element $x \in X$ such that $f(x) = x$. ``Fixpoint'' is (standard) jargon for fixed point. For some historical information on this theorem and its variations, see \cite{LNS}.}

\begin{theorem}[The Fixpoint Theorem]
Let $D$ be an $\omega$-cpo with a least element, and $f : D \rightarrow D$ a continuous function. Then $f$ has a least fixed point $\lfp{f}$. Moreover, $\lfp{f}$ is defined explicitly by:
\begin{equation}
\label{lfpeq}
 \lfp{f} = \olub f^{n} (\bot ) . 
\end{equation}

\end{theorem}

We give the proof, since it is elementary, and exhibits very nicely how the basic axiomatic structure of Domains is used.

\begin{proof}
Note that $f^{n}(\bot)$ is defined inductively by:
\[ f^{0}(\bot) = \bot,  \qquad f^{k+1}(\bot) = f(f^{k}(\bot)) . \]
We show firstly that this sequence is indeed an \ochain. More precisely, we show for all $k \in \Nat$ that $f^{k}(\bot) \sqle f^{k+1}(\bot)$. For $k = 0$, this is just $\bot \sqle f(\bot)$. For the inductive case, assume that $f^{k}(\bot) \sqle f^{k+1}(\bot)$. Then by monotonicity of $f$, $f(f^{k}(\bot)) \sqle f(f^{k+1}(\bot))$, \ie $f^{k+1}(\bot) \sqle f^{k+2}(\bot)$, as required.

Next we show that (\ref{lfpeq}) does yield a fixpoint. This is a simple calculation using the continuity of $f$:
\[ f( \olub f^{n} (\bot ) ) = \olub f^{n+1} (\bot) = \olub f^{n} (\bot ) . \]
The last step uses the (easily verified) fact that removing the first element of an \ochain  does not change its least upper bound.

Finally, suppose that $a$ is a fixpoint of $f$. Then we show by induction that, for all $k$, $f^{k}(\bot) \sqle a$. The basis is just $\bot \sqle a$. For the inductive step, assume $f^{k}(\bot) \sqle a$. Then by monotonicity of $f$, 
\[ f^{k+1}(\bot) = f(f^{k}(\bot)) \sqle f(a) = a . \]
Thus $a$ is an upper bound of $(f^{n}(\bot) \mid n \in \omega)$, and hence $\olub f^{n} (\bot ) \sqle a$.
\end{proof}

\paragraph{Factorial revisited}
We now reconstrue the definition of the factorial function  we considered  previously,  as a function on \emph{domains}:
\[ F : \Pfn{\Nat}{\Nat} \longrightarrow \Pfn{\Nat}{\Nat} , \]
defined by
\[ F(f)(n) =  \cond{n=0}{1} {n \times f (n-1)} . \]
We can check that \emph{$F$ is continuous}.  Hence we can apply the fixpoint theorem to $F$, and conclude that it has a least fixpoint $\lfp{f}$, defined explicitly by (\ref{lfpeq}). Now we can make the (explicit, non-circular) definition:
\[ \fact = \lfp{F} . \]
One can check   that this definition yields exactly the expected definition of factorial. In fact, the increasing sequence constructed in forming the least fixpoint according to (\ref{lfpeq}) is exactly the one we described concretely in our previous discussion of the factorial.

Thus in particular the processes of information increase we have been emphasizing are involved directly in the construction underpinning the Fixpoint Theorem.

\subsubsection{Further Developments in Domain Theory}
This is of course just the beginning of an extensive subject. We  mention a few principal further features of Domain Theory:
\begin{description}
\item[Function Spaces] 
A key point of the theory is that, given domains $D$ and $E$,   the set of continuous functions from $D$ to $E$, written as $\funsp{D}{E}$, will again be a domain, with  the following \emph{pointwise ordering}:
\[ f \sqle g \;\; \Longleftrightarrow \;\; \forall x \in D. \, f(x) \sqle_{E} g(x) . \]
Moreover, operations such as function application and currying or lambda-abstraction are continuous. This means that we can form models of typed $\lambda$-calculi and higher-order computation within Domain Theory, which is of central importance for the  denotational semantics of programming languages. Of course, such domains of higher-order functions are very ``abstract''---they are in fact the prime examples of domains which are \emph{not} concrete in the sense of \cite{KP78}---and notions of temporality are left quite far behind. (There have attempts to capture more of these notions by varying the definition of the order on function spaces, but these have not been completely successful---and in some cases, provably cannot be).
\item[Recursive Types]
Remarkably, the idea of the Fixpoint Theorem, and its use to give meaning to recursive definitions of elements of domains, can be lifted to the level of domains themselves, to give meaning to \emph{recursive definitions of types}. This even extends to the free use of function spaces in recursive definitions of domains, leading to the construction of domains $D$ whose continuous function spaces $\funsp{D}{D}$ are isomorphic to $D$ or to a  subspace of $D$. This allows models of the type-free $\lambda$-calculus, and of various strongly impredicative type theories, to be given within Domain Theory.
\item[Powerdomains]
There are also a number of  \emph{powerdomain} constructions $P(D)$, which build a domain of subsets of $D$. This allows various forms of non-deterministic and concurrent computation to be described. There is also a \emph{probabilistic powerdomain} construction, which provides semantics for probabilistic computation.
\end{description}

\paragraph{Some suggestions for further reading on Domain Theory}
The text \cite{DP02} gives a fairly gentle introduction to partial orders and lattices, with some material on domains. The Handbook article \cite{AJ} is a comprehensive technical survey of domain theory. The monograph \cite{compendium} focusses on the connections to topology and lattice theory. Gordon Plotkin's classic lecture notes \cite{Plo} are available on-line. The texts \cite{Win,AmCu} show how domain theory is used in the semantics of programming languages.

\subsection{Dynamic Logic}
Dynamic Logic originates at the confluence of two sources: modal logic and its Kripke semantics \cite{Kri63,BRV}; and Hoare logic of programs \cite{Hoa69}.
\paragraph{Modal Logic}
Modal Logic adds to a standard background logic (say classical propositional calculus) the propositional operators $\Box$ and $\Diamond$, expressing ideas of  ``necessity'' and ``possibility''.  This was transformed from a philosophical curiosity to a vibrant and highly applicable branch of mathematical logic by the introduction of Kripke semantics \cite{Kri63}. This is based on Kripke structures $(W, R, V)$, where $W$ is a set of worlds, $R \subseteq W \times W$ is an ``accessibility relation'', and $V : \Prop \rarr \Pow{W}$ is a valuation which for each propositional atom in $\Prop$ assigns the set of worlds in which it is true. This valuation is then extended to one on formulas, with the key clauses:
\[ w \models \Box \phi \;\; \equiv \;\; \forall w'. \, wRw' \; \Rightarrow \; w' \models \phi  \]
\[ w \models \Diamond \phi \;\; \equiv \;\; \exists w'. \, wRw' \; \wedge \; w' \models \phi . \]
The importance of the Kripke semantics is that it gives modal logic a clear mathematical purpose: it is a logical language for talking about such structures, which strikes a good balance between expressive power and tractability. Computer Science provides a wealth of situations where such structures arise naturally, and where there is a clear need for the verification of their logical properties. 
The dominant interpretation of Kripke structures in Computer Science replaces metaphysical talk of  ``possible worlds'' by the more prosaic terminology of \emph{states}. Here we think of  states of a system, which are generally characterized by the information we have about them. In a Kripke structure, the \emph{direct} information we have about a state is which atomic propositions are true in that state. However, while we seem again to be speaking about information states, as in our discussion of Domain Theory, there is an important difference. In Domain Theory, (as in Kripke semantics for Intuitionistic Logic), information is in general \emph{partial}, but also \emph{persistent}. Information can only \emph{increase} along a computation. We may never reach total information, but we will never lose what we had---just as we can never (in current Physics) change the past. (Indeed, the two are intimately related. In the implicit temporality of Domain Theory, the current information state summarizes all the information produced in the computation up till now; whatever happens in the future cannot change that). By contrast, Kripke structures for modal logics correspond to a less stable world. We may have perfect knowledge of the current state, but the dynamics of the system, as described by the accessibility relation, allow in general for arbitrary state change. A basic Computer Science model for this scenario is provided by taking the states to be memory states of a computer. At some instant of time we may have a complete snap-shot of the memory. But our repertoire of actions allow us to assign an arbitrary new value into any memory cell, so we can go from any given state to any other (possibly by a sequence of basic actions).
In particular, the key feature of computer memory, the fact that we can destructively over-write the previous contents of a memory cell, (a feature which is not, apparently, available for our own memories!), ensures that the past is not in general carried forward. 

\paragraph{Hoare Logic}
Hoare Logic \cite{Hoa69,deBa} provides a compositional proof theory for reasoning about imperative programs. It is a two-sorted system. We have a syntax for \emph{programs} $P$, and one for \emph{formulas} $\phi$, which are generally taken to be formulas of predicate calculus. Such formulas can be used to express properties of program states (\ie memory state snap-shots as in our previous discussion, or more formally assignments of values to the variables appearing in the program), by a~\textbf{ variable pun} by which the individual variables used in formulas are identified with the program variables. The basic assertions of the system are taken to be \emph{Hoare triples} $\phi\{P\}\psi$. Such a triple is said to be valid if, in any initial state satisfying the formula $\phi$ (the \emph{precondition}), execution of the program $P$ will, if it terminates, result in a final state satisfying the formula $\psi$ (the \emph{post-condition}). 

The \textbf{variable pun} is put to use in the axiom for assignment statements:
\[ \phi[e/x] \{x := e\} \phi \]
which says that $\phi$ is true after executing the assignment statement $x := e$ if $\phi$ with $e$ substituted for $x$ was true before.

The key rules of the system allow for compositional derivation of assertions about complex programs from assertions about their immediate sub-programs.
\[
\frac{\phi\{P\}\psi \quad \psi\{Q\}\theta}{\phi\{P;Q\}\theta} \qquad
\frac{\phi \wedge B\{P\}\psi \quad \phi \wedge \neg B\{Q\}\psi}{\phi\{ \cond{B}{P} {Q} \}\psi} \qquad
\frac{\phi \wedge  B\{P\}\phi}{\phi \{\while{B}{P}\}\phi \wedge \neg B}
\]
Here $P ;Q$ is the \emph{sequential composition} which firstly performs $P$, then $Q$; $\cond{B}{P} {Q}$ is the \emph{conditional} which evaluates $B$ in the current state; if it is \textbf{true} then $P$ is performed, while if it is \textbf{false}, $Q$ is performed. Finally, $\while{B}{P}$ evaluates $B$; if it is \textbf{true}, then $P$ is performed, after which the whole statement is repeated; while if it is \textbf{false}, the statement terminates immediately.

\paragraph{Dynamic Logic}
Dynamic Logic \cite{Pra76} arises by combining salient features of these two systems. Note that we are reasoning about programs in terms of the \emph{input-output relations on states} which they define. If the program is deterministic, this relation will actually be a partial function, but there is no need to insist on this. We can thus view each program $P$ as defining a relation $R \subseteq S \times S$, where $S$ is the set of states. Thus for each individual program, we obtain a Kripke structure $(S, R, V)$, where $V$ is the valuation which assigns truth conditions on states for some repertoire of state predicates. The key point of contact between the two systems is that validity of the Hoare triple $\phi\{P\}\psi$ corresponds exactly to the validity of the modal formula
\[ \phi \rarr \Box \psi \]
in the Kripke structure $(S, R, V)$, where by validity we mean that
\[ s \models \phi \rarr \Box \psi \]
for every $s \in S$. 

As a first extension, we can consider multiple programs, each defining an accessibility relation $R$. To keep track of which program we are talking about at any given point, we replace $\Box$ by $[R]$, so that the formula corresponding to the Hoare triple now reads as
\[ \phi \rarr [R] \psi . \]
Just as $[R]$ replaces $\Box$, so $\langle R \rangle$ replaces $\Diamond$. Thinking of $R$ as the input-output relation defined by a program, we can read $[R]\phi$ as holding in all states (worlds) $s$ such that any output state obtained by executing $R$ starting in $s$ will satisfy $\phi$. Similarly, $\langle R \rangle$ will be true in any state $s$ such that there is some output state than can be obtained by running $R$ starting in $s$ which satisfies $\phi$.

This is just \emph{multi-modal logic}, with mutiple accessibility relations, each with its own modalities. Note that it is now completely meaningful to consider modal formulas which make assertions about programs which go well beyond Hoare triples, e.g.
\[ [R]\langle S \rangle \phi \rarr \langle S \rangle [R] \phi . \]
However, at this point we lack the compositional analysis of programs offered by Hoare Logic.

The final step to (propositional) Dynamic Logic comes by considering a two-sorted system with a mutually recursive syntax. We have a set $\Prop$ of propositional atoms as before, and also a set $\RV$ of basic relations. The syntax of formulas is given by
\[ \phi \;\; ::= \;\; p \in \Prop \mid \neg \phi \mid \phi \; \wedge \; \psi \mid [R]\phi \]
while the syntax of relations $R$ is given by
\[ R \;\; ::= \;\;  r \in \RV \mid R;S \mid R \cup S \mid R^{\ast} \mid \phi? \]
We have not included the modal operator $\langle R \rangle$ as primitive syntax, since we can define
\[ \langle R \rangle \phi \;\; \equiv \;\; \neg [R] \neg \phi . \]

In this syntax, any program is allowed to appear as a modal operator on formulas, while in addition to the usual regular operations of relational algebra (composition, union, and reflexive transitive closure), any formula is allowed to appeared as a program test (we may call this the \textbf{formula pun}). In general, this is too strong, and only a restricted class of tests should be allowed. Tests are interpreted as sub-identity relations---so $\phi?$ is the set of all $(s, s)$ such that $\phi$ is true in $s$.

Note that the usual imperative program constructs can be recovered from these relational constructs. Sequential composition is provided directly, while
\[ \cond{b}{R}{S} \; \equiv \; b;R \cup \neg b;S \qquad \while{b}{R} \; \equiv \; (b;R)^{\ast};\neg b . \]
The Hoare Logic axioms can now be derived from the following modal axioms:
\[
\begin{array}{rcl}
[R;S]\phi  & \leftrightarrow &   [R][S]\phi \\
\DLbox{R\cup S}\phi   & \leftrightarrow   & [R]\phi \; \wedge \; [S]\phi \\
\DLbox{\psi?}\phi  & \leftrightarrow & \psi \rarr \phi 
\end{array}
\]
and the rule
\[ \frac{\phi \rarr [R]\phi}{\phi \rarr [R^{\ast}]\phi} . \]

\subsubsection{Discussion}

While Hoare Logic is specifically tailored to the needs of conventional imperative programming languages, Dynamic Logic is much more generic in style; and indeed, it has been applied in a range of contexts, including Natural Language and Quantum Logic. In the Chapter in this Handbook by Baltag and Moss, a version of Dynamic Logic is described in which the states are \emph{information states of agents}, and the actions are \emph{epistemic actions} by these agents, such as public announcements.

As a general formalism, though, Dynamic Logic offers only a limited analysis of information dynamics. Indeed, despite its name, it is not really very dynamic, as it is limited to speaking  of the input-output behaviour of relations. This is confirmed by the simple translation it admits into first-order logic (augmented with fixpoints to account for the reflexive transitive closure operation on relations).
We briefly sketch this. To each relation term $R$, we associate a formula $B_{R}(x, y)$ in two free variables, and to each modal formula $\phi$ we associate a formula $A_{\phi}(x)$ in one free variable.
The main clauses in the definition of $B_{R}$ are as follows:
\[ \begin{array}{lcl}
B_{R \cup S}(x, y) & \equiv & B_{R}(x, y) \; \vee \; B_{S}(x, y) \\
B_{R ; S}(x, y) & \equiv & \exists z. \, B_{R}(x, z) \; \wedge \; B_{S}(z, y) \\
B_{\phi ?}(x, y) & \equiv & A_{\phi}(x) \; \wedge \; x = y \\
B_{R^{*}}(x, y) & \equiv & \mu S. \, [(x = y) \; \vee \; (\exists z. \, B_{R}(x, z) \, \wedge \, S(z, y))]
\end{array}
\]
The clauses for modal formulas are standard. The one for the modality is:
\[ A_{[R]\phi}(x) \; \equiv \; \forall y. \, B_{R}(x, y) \rarr A_{\phi}(y) . \]

\paragraph{Suggestion for further reading}
The book \cite{HKT} is a comprehensive technical reference, while \cite{JvBbook} is a wide-ranging study. Applications to Natural Language appear in \cite{GrSt,vESt}, and to Quantum Logic in \cite{BS06}.

\subsection{Process Algebra}
\subsubsection{Background}

One of the major areas of activity in Theoretical Computer Science over the past three decades has been Concurrency Theory, and in particular Process Algebra. Whereas  modelling  sequential computation in terms of input-output functions or relations essentially uses off-the-shelf tools from Discrete Mathematics and Logic, albeit in novel combinations and with new technical twists, and even Domain Theory can be seen as an off-shoot of General Topology and Lattice Theory, Concurrency Theory has really opened up some new territory. In Concurrency Theory, the computational processes themselves become the objects of study; concurrent systems are executed for the behaviour they produce, rather than to compute some pre-specified function. In this setting, even such corner-stones of computation as Turing's analysis of computability do not provide all the answers. For all its conceptual depth, Turing's analysis of computability was still calibrated using familiar mathematical objects: which \emph{functions} or \emph{numbers} are computable? When we enter the vast range of possibilities for the behaviour of computational systems in general, the whole issue of what it means for a concurrent formalism to be \emph{expressively complete} must be re-examined. There is in fact no generally accepted form of Church-Turing thesis for concurrency; and no widely accepted candidate for a universally expressive formalism. Instead, there are a huge range of concurrency formalisms, embodying a host of computational features.

Another question which ramifies alarmingly in this context is what is the right notion of \emph{behavioural equivalence} of processes. Again, a large number of candidates have arisen. Experts use what seems most appropriate for their purpose; it is not even plausible that a single notion will gain general acceptance as ``the right one''.

In fact, a great deal of progress has been achieved, and the situation is much more positive than might appear from these remarks. There is a great diversity of particular formalisms and definitions in Concurrency Theory; but underpinning these are a much smaller number of underlying paradigms and technical tool-kits, which do provide effective intellectual instruments, both for fundamental research and applications.

Examples include: 
\begin{itemize}
\item labelled
transition systems and bisimulation
\item naming and scope restriction and
extrusion
\item 
the automata-theoretic paradigm for model-checking
\end{itemize}
These tool-kits are the real fruits of these theories. They may be
compared to the traditional tool-kits of physics and engineering:
Differential Equations, Laplace and Fourier Transforms, Numerical 
Linear Algebra, etc.
They can be applied to a wide range of situations, going well beyond those originally envisaged, e.g.
Security, Computational Biology, and Quantum Computation.

\subsubsection{Some Basics of Process Algebra}
We now turn to a brief description of a few basic notions, in a subject on which there is a vast literature.
We begin with the key semantic structure, namely \emph{labelled transition systems}.
A labelled transition system is a structure $(S, \Act, T)$, where $S$ is a set of states, $\Act$ is a set of actions, and $T \subseteq S \times \Act \times S$ is the transition relation. We write $\trans{s}{a}{t}$ for $(s, a, t) \in T$. Note how close this is to the notion of Kripke structure we have already encountered. However, that notion is tuned to a \emph{state-based} view of computation, in which we focus on assertions which are true in given states. The transition relation plays an indirect r\^ole, in controlling the behaviour of the modal operators. By contrast, the point of view in labelled transition systems is that states are not directly observable, and hence do not have properties directly attributable to them. Rather, it is the actions which are the basic observables, and we infer information about states indirectly from  their potential for observable behaviour. Thus the point of view here is closer to automata theory. A key difference from classical automata theory, however, is that we look beyond the classical notion of behaviour in terms of the words or traces (sequences of actions) accepted or generated by the system, and also encompass \emph{branching behaviour}. The classical example which illustrates this is the following \cite{Mil80}:
\[ 
\begin{diagram}[2em]
& & \bullet & & \\
& & \dTo^{a} & & \\
 & & \bullet & & \\
& \ldTo^{b} & & \rdTo^{c} & \\
\bullet & & & & \bullet
\end{diagram}
\qquad \qquad \not\sim \qquad \qquad
\begin{diagram}[2em]
& & \bullet & & \\
& \ldTo^{a} & & \rdTo^{a} & \\
\bullet & & & & \bullet \\
\dTo^{b} & & & & \dTo_{c} \\
\bullet & & & & \bullet \\
\end{diagram}
\]
These systems have the same linear traces $\{ ab, ac \}$. However,  if we think of a scenario where we can perform  experiments by  pressing buttons labelled with the various actions, and observe if the  experiments succeed, \ie whether the system performs the corresponding action, then after observing an $a$ in the first system, it is clear that whether we press the $b$ button or the $c$ button, we will succeed; whereas in the second system, one button will succeed and the other won't. A fundamental notion of process equivalence which enforces this distinction is \emph{bisimulation}.
We define a \emph{bisimulation} \cite{JvB76,HM80,Par81,Mil89,San04} on a  labelled transition system $(S, \Act, T)$ to be a relation $R \subseteq S \times S$ such that:
\[ \begin{array}{lcl}
s R t \; \wedge \; \trans{s}{a}{s'} & \Rightarrow & \exists t'. \, \trans{t}{a}{t'} \; \wedge \; s'Rt' \\
\wedge & & \\
s R t \; \wedge \; \trans{t}{a}{t'} & \Rightarrow & \exists s'. \, \trans{s}{a}{s'} \; \wedge \; s'Rt'
\end{array}
\]
We write $s \sim t$ if there is a bisimulation $R$ such that $s R t$. We can see that  indeed the root states of the two trees in the above example are not bisimilar, since the first has an action $a$ to a state in which both the actions $b$ and $c$ are possible, while the second has no $a$-move to a matching state.

We now turn to a suitable modal logic for labelled transition systems. The basic form for such a logic is Hennessy-Milner Logic. This has modal operators $[a]$, $\langle a \rangle$ for each action $a$. In general, this logic does not have (or require) any propositional atoms; just constants $\true$ (true) and $\false$ (false).
The semantic clauses are as expected for a multi-modal logic, where we view the transition relation as an $\Act$-indexed family of relations $\{ T_{a} \}_{a \in \Act}$, where $T_{a} \subseteq S \times S$ is defined by
\[ T_{a} = \{ (s, t) \mid (s, a, t) \in T \} . \]
Thus we have the clauses
\[ s \models [a]\phi \;\; \equiv \;\; \forall t. \, \trans{s}{a}{t} \; \Rightarrow \; t \models \phi \]
\[ s \models \langle a \rangle\phi \;\; \equiv \;\; \exists t. \, \trans{s}{a}{t} \; \wedge \; t \models \phi . \]
The basic result here is that, under suitable hypotheses, two states in a labelled transition system are bisimilar if and only if they satisfy the same formulas in this modal logic. Thus in our example above, the first system satisfies the formula $\langle a\rangle(\langle b \rangle\true \; \wedge \; \langle c \rangle\true)$, while the second does not.

We now turn, finally, to the \emph{algebraic} aspect of process algebra. Just as we structured the programs in  Dynamic Logic using relational algebra, so we seek an algebraic structure to generate a wide class of process behaviours. As we have already discussed, there is no one universally adopted set of process combinators, but we shall consider a standard set of operations, essentially a fragment of Milner's CCS \cite{Mil80,Mil89}.
The syntax of process terms $P$ is defined, assuming a set $\Act$ of actions,  as follows:
\[
P \;\; ::= \;\; a.P \;\; (a \in \Act) \mid P + Q \mid \; 0 \; \mid P \parallel Q . \]
Here $a.P$ is \emph{action prefixing}; first do $a$, then behave as $P$. $P + Q$ is \emph{non-deterministic choice} between $P$ and $A$, while $0$ is \emph{inaction}; the process which can do nothing. Finally, $P \parallel Q$ is \emph{parallel composition}, which we take here in a  simple form, not involving any interaction between $P$ and $Q$.

We formalize these intuitions as a labelled transition system in which the states are the process terms, while the transition relation is defined by structural induction on the syntax of terms---the Structural Operational Semantics paradigm \cite{PloSOS}.

The transition relation is specified as follows.
\[
\frac{}{\trans{a.P}{a}{P}} \qquad \frac{\trans{P}{a}{P'}}{\trans{P+Q}{a}{P'}} \qquad 
 \frac{\trans{Q}{a}{Q'}}{\trans{P+Q}{a}{Q'}} 
 \]
 \[ \frac{\trans{P}{a}{P'}}{\trans{P\parallel Q}{a}{P' \parallel Q}} \qquad
 \frac{\trans{Q}{a}{Q'}}{\trans{P\parallel Q}{a}{P \parallel Q'}}
 \]
 This labelled transition system gives rise to a notion of bisimulation, which is an equivalence relation, and in fact a congruence for the process algebra. The corresponding equational theory for the algebra can be axiomatized as follows:
 \[ \begin{array}{rcl}
 P + P & = & P \\
 P + 0 & = & 0 \\
 P + Q & = & Q + P\\
 P + (Q + R) & = & (P + Q) + R \\
 \end{array}
 \]
 together with the following equational scheme. If $P \equiv \sum_{i \in I} a_{i}.P_{i}$ and $Q \equiv \sum_{j \in J} b_{j}.Q_{j}$, then:
 \[ P \parallel Q \;\; = \;\; \sum_{i \in I} a_{i}.(P_{i} \parallel Q) + \sum_{j \in J} b_{j}.(P \parallel Q_{j}) . \]
This is an infinite family of equations.  In fact, the equational theory of bisimulation on process terms is not finitely axiomatizable \cite{Mol90b}; however, with the aid of an auxiliary operator (the ``left merge''), a finite axiomatization can be achieved \cite{Mol90a}.

\subsubsection{Communication and Interaction in Process Algebra}
We shall take a brief glimpse at this large topic. For illustration, we shall describe the CCS approach \cite{Mil80}. However, it should be emphasized that there is a huge diversity of approaches in the process algebra literature, with none having a clear claim to being considered canonical. (For further remarks on this issue of non-canonicity, see the final section of this article, and \cite{Abr06}).

We assume some structure on the set $A$: a fixed-point free involution $a \mapsto \bar{a}$, so that we have
$a \neq \bar{a}$ and $\bar{\bar{a}} = a$. The idea is that $a$ and $\bar{a}$ will be complementary partners to an interaction or synchronized action. We also introduce a special action $\tau$, which is intended to be a ``silent action'', unobservable to the external environment.

We can now introduce a parallel composition $P \spar Q$ which does allow for interaction, in the form of synchronization between $P$ and $Q$. Its dynamics are given by the following rules:
\[ \frac{\trans{P}{a}{P'}}{\trans{P \spar Q}{a}{P'  \spar  Q}} \qquad
\frac{\trans{P}{a}{P'} \quad \trans{Q}{\bar{a}}{Q'}}{\trans{P \spar Q}{\tau}{P' \spar Q'}} \;\; (a, \bar{a} \neq \tau) \qquad
 \frac{\trans{Q}{a}{Q'}}{\trans{P \spar Q}{a}{P \spar Q'}}
 \]
The new ingredient is the middle rule, which allows for synchronization between $P$ and $Q$. Note that $a$ and $\bar{a}$ ``complete'' each other into the action $\tau$ which is now an internal step of the system, and hence unobservable to the external environment.
 
 To take proper account of the unobservable character of $\tau$, we introduce the observable transition relation for each $a \neq \tau$:
 \[ \otrans{}{a}{} \;\; = \;\; {\trans{}{\tau}{}}^{*}  \trans{}{a}{} {\trans{}{\tau}{}}^{*} \]
 We can then define \emph{weak bisimulation} with respect to these observable transitions. However, a new complication arises: this weak bisimulation is not a congruence with respect to the operations of the process algebra. It is necessary to take the largest congruence compatible with weak bisimulation, finally yielding the notion of \emph{observational congruence}. This notion can be equationally axiomatized, but it is considerably more complex and less intuitive than the ``strong bisimulation'' we encountered previously.
 
\subsubsection{Discussion}
Process Algebra can be used as a \emph{vehicle} for discussions of information flow and information dynamics, e.g.~\cite{Lowe02}. It does not in itself offer a fully fledged theory of these notions.

Process Algebra is a qualitative theory of process behaviours. It is our first example of a 
\emph{dynamic} theory, since it makes temporality and the flow of events explicit.

\paragraph{Suggestions for further reading}
Introductory textbooks include \cite{HoareCSP,Mil89,BaWe90,Mil99}. The Handbook of Process Algebra \cite{HandProcAlg}
provides wide technical coverage of the field.

\section{Combining Qualitative and Quantitative Theories of Information}

\subsection{Scott domain theory and Shannon information theory}
Two important theories of information give contrasting views on the question of information increase, which we discussed in Section~1. Information theory \textit{\`a la} Shannon is a quantitative theory which considers how \emph{given} information can be transmitted losslessly on noisy channels. In this process, information may only be lost, never increased.
Domain Theory \textit{\`a la} Scott is, as we have seen, a qualitative theory in which the key notion is the partial order $x \sqsubseteq y$, which can be interpreted as: ``$y$ has more information content than $x$''. This theory is able to model a wide range of computational phenomena. To take a classical example, consider the interval bisection methods for finding the root of a function. We start with an interval in which the root is known to lie. At each step, we halve the length of the interval being considered. This represents an increase in our information about the location of the root, in an entirely natural sense. In the limit, this nested sequence of intervals contains a single point, the root -- we have perfect information about the solution.

More generally, in Domain Theory recursion (and thereby control mechanisms such as iteration) is modelled as the least fixed point of a monotonic and (order-)continuous function:
\[ \bot \sqsubseteq f \bot \sqsubseteq f^2 \bot \sqsubseteq \cdots \bigsqcup_{k} f^k \bot \]
since $f(\bigsqcup_{k} f^k \bot ) = \bigsqcup_{k} f^{k+1} \bot = \bigsqcup_{k} f^k \bot$.
Thus a basic tenet of this theory is that information \emph{does} increase during computation, and in particular this is how the meaning of recursive definitions is given.

It is intriguing to consider that the different viewpoints taken by Information theory and Domain Theory may have been influenced by their technological roots.
Information theory was summoned forth by the needs of the telecommunications industry, whose task is to transmit the customer's data with the highest  possible fidelity. Domain Theory arose as a mathematical theory of \emph{computation}; the task of computation is to ``add value'' to the customer's data.\footnote{Which of course raises our question of how this can be possible thermodynamically. The answer is, again, that it is the \emph{customer's data} which is having value added to it; just as buying energy from the National Grid does not violate the Second Law.}

How can these views be reconciled? Information theory is a thermodynamic theory; Shannon information is negative entropy. From this viewpoint, the \emph{total information} of a system can only decrease; however, information can flow from one subsystem into another, just as a body can be warmed by transferring heat from its environment.

The Domain Theory view, we suggest, arises most naturally if we think of adding an observer to a system. It is the observer's information which increases during a computation. This reading has a precise mathematical analogue in the view of Domain Theory as a ``logic of observable properties'' \cite{Abramsky91}.
Information increase is always, necessarily it seems, \emph{relative to a sub-system}. Moreover, this is a  subsystem which can \emph{observe} its environment, and which may, symmetrically, act itself to direct information to the environment. It is then a small step to viewing such sub-systems as \emph{agents}.

It is worth adding that Shannon Information Theory also relies on such a view for its guiding intuitions. One of the standard ways of motivating Shannon information is in terms of ``twenty questions'': the number of bits of information in a message is how many yes/no questions we would need to have answered in order to know the contents of the message. Again, implicit here is some interaction between agents. And of course, the purpose of communication itself is to transfer information from one agent to another.

We need a quantitative theory to deal with essentially quantitative
issues such as complexity, information content, rate of information
flow etc. However, the \emph{weakness} of a purely quantitative theory
is that numbers are always comparable, so that some more subtle issues are
obscured, such as, crucially, distinguishing different \emph{directions} of
information increase. Beyond this, by combining quantitative and
qualitative aspects, e.g. in formulating conditions on ``informatic
processes'', a unified theory can be more than the sum of its parts.

\subsection{Domains with measurements: connecting the quantitative and
  qualitative views}
An important step towards unifying the qualitative and the quantitative
points of view was taken in Keye Martin's Ph.D. thesis \cite{KeyeThesis} and
  subsequent publications \cite{Mar2,Mar3,MMW}. Martin introduced a simple notion of
  \emph{measurement} on domains. In its most concrete form, a
  measurement assigns real numbers to domain elements, which can be
  said to measure the degree of undefinedness or uncertainty of the
  element. Thus the maximal elements, which can be regarded as having
  perfect information, will have measurement 0. The axioms for
  measurements, while quite simple and intuitive, tie the quantitative 
  notion in with the qualitative domain structure in a very rich
  way. Just to mention some of the highlights:
\begin{itemize}
\item There is a rich theory of fixpoints which applies to increasing, 
  \emph{but not necessarily monotonic}, functions on domains. This is
  already a remarkable departure from `classical' domain theory, in
  which monotonicity is always assumed. However, Martin shows that
  there are compelling natural examples, such as interval bisection,
  which require this broader framework. Not only are there existence
  and uniqueness
  theorems for fixpoints in this frameworks, but also novel induction principles.
\item As the previous point suggests, there is a move away from the
  use of domain theory to model purely extensional aspects of
  computation, and towards using it to capture important features of
  \emph{computational processes}. This leads to a notion of `informatic
  derivative' which can be used to gain information about the \emph{rate of
  convergence} of a computational process.
\item A notable aspect of this development is the unified basis on
  which it puts the study of both discrete and continuous
  (e.g. real-number) computation. 
\end{itemize}
It is also important that there are many natural examples of
measurement covering most of the domains standardly arising as
data-types for computation, including the domain of intervals, for which the natural measurement is the \emph{length} of the interval; finite lists and other standard finite
data-structures; streams; partial functions on the natural numbers;
and both non-deterministic and probabilistic powerdomains.

However, the example which has really revealed the possibilities of
this framework has only appeared recently, and is a major development
in its own right.

\subsection{Combining Scott Information and Shannon Information}
Recently, Bob Coecke and Keye Martin have produced a very interesting construction which can be seen as a first step towards a unification of these two theories of information \cite{Coecke2002}.
The problem which they attacked can be formulated as follows.
Consider the set of probability distributions on a finite set. For an
$n$-element set, these are the ``classical $n$-states'' of Physics:
\[\Delta^n:=\{ x\in[0,1]^n:\sum_{i=1}^nx_i=1\} .\]
This is 
the setting in which Shannon entropy, the fundamental quantitative
notion in classical Information Theory, is defined. It assigns a
number, the ``expected information'', to each classical state. The question is: can we place a \emph{partial order} on
$\Delta^n$ such that:
\begin{enumerate}
\item This partial order forms a domain.
\item Shannon entropy is a measurement with respect to this domain.
\item The order extends to quantum states (density operators).
\end{enumerate}
These are highly non-trivial requirements to satisfy. Note that the
set of
probability distributions on a 3-element set, seen as a subset of
Euclidean space,  form a (solid) triangle, 
and in general those on a $n$-element set form an $n$-simplex.
The distribution corresponding to maximum uncertainty is the uniform distribution, with
each point assigned probability $1/n$ --- geometrically, the
barycenter of the simplex; while the maximal elements $\mathrm{max}(\Delta^n)$, corresponding
to perfect information,  are the \emph{pure states} assigning probability 1
to one element, and 0 to all others --- geometrically, the vertices
of the simplex. This geometrical aspect brings a rich mathematical
structure to this example which seems different to anything previously 
encountered in Domain Theory.

Note also the contrast with previous work on the
probabilistic powerdomain \cite{Jones}. Classical probability distributions are
\emph{maximal elements} in the probabilistic powerdomain; non-standard
elements (valuations) are introduced which provide approximations to
measures, but the order restricted to the measures themselves is
discrete.
By contrast, we are seeking a rich informatic structure on the
standard objects of probability (distributions) and quantum mechanics
(density operators) \emph{themselves}, without introducing any non-standard
elements. It is by no means \textit{a priori} obvious that this can be 
done at all; once we see that it can, many new possibilities will unfold.

A classical state $x\in\Delta^n$ is \em pure \em when $x_i=1$ for some
$i\in\{1,\ldots,n\}$; we denote such a state by $e_i$.
Pure states $\{e_i\}_i$ are the actual states a system can 
be in, while general mixed states $x$ and $y$ are epistemic entities. 

If we know $x\in\Delta^{n+1}$ and
by some means determine that outcome $i$ is
not possible, our knowledge
improves to
\[p_i(x)=\frac{1}{1-x_i}(x_1,\ldots,\hat{x_i},\ldots,x_{n+1})\in\Delta^n,\]
where $p_i(x)$ is obtained
by first removing $x_i$ from $x$ and
then renormalizing. The partial mappings which result,
\[p_i:\Delta^{n+1}\rightharpoonup\Delta^n\]
with dom$(p_i)=\Delta^{n+1}\setminus\{e_i\}$,
are called the {\it Bayesian projections\,} and lead
one directly to the following inductively defined relation on classical states.

\begin{definition}
For $x,y\in\Delta^2$:
\begin{equation}\label{twostates}
x\sqsubseteq y\equiv (y_1\leq x_1\leq 1/2)\mbox{ or }(1/2\leq x_1\leq
y_1)\,.
\end{equation}
\em For  $n \geq 2$, and $x,y\in\Delta^{n+1}$:
\begin{equation}\label{inductiverule}
x\sqsubseteq y\equiv(\forall
i)(x,y\in\mbox{dom}(p_i)\Rightarrow p_i(x)\sqsubseteq p_i(y)).
\end{equation}
The relation $\sqsubseteq$ on $\Delta^n$ is called the \em Bayesian
order. \em
\end{definition}
See \cite{Coecke2002} for motivation, and results showing that the order on
$\Delta^2$ is uniquely determined under minimal assumptions.

The key result is:
\begin{theorem}
$(\Delta^n,\sqsubseteq)$ is a domain with
maximal elements
\[\mathrm{max}(\Delta^n)=\{e_i:1\leq i\leq n\}\]
and least element $\bot:=(1/n,\ldots,1/n)$. Moreover,
Shannon entropy
\[\mu (x) =-\sum_{i=1}^nx_i\log x_i\]
is a
measurement of type $\,\Delta^n\rightarrow[0,\infty)^*.$
\end{theorem}

The Bayesian order can also be described in a
more direct manner, the \em symmetric characterization. \em
Let $S(n)$ denote the group of permutations on $\{1,\ldots,n\}$,
and 
\[ {\Lambda^n:=\{x\in\Delta^n:(\forall i<n)\,x_i\geq x_{i+1}\}} \]
the collection of \em monotone \em
classical states.

\begin{theorem}
\label{classicalsymmetries}
For $x,y\in\Delta^n$, we have $x\sqsubseteq y$ iff
there is a permutation ${\sigma\in S(n)}$ such that
$x\cdot\sigma,y\cdot\sigma\in\Lambda^n$
and
\[(x\cdot\sigma)_i(y\cdot\sigma)_{i+1}\leq
(x\cdot\sigma)_{i+1}(y\cdot\sigma)_i\]
for all $i$ with $1\leq i<n$.
\end{theorem}
In words, this result says that the Bayesian order holds between states $x$ and $y$ if we can find a permutation $\sigma$ which rearranges them both as monotone states, and such that $x$ falls less rapidly than $y$ as we proceed through the ordered list of component probabilities.

Thus, the Bayesian order
is order isomorphic to $n!$ many copies of $\Lambda^n$
identified along their common boundaries. This fact,
together with the pictures of ${\uparrow} x := \{ y \in \Delta^{n} \mid x \sqsubseteq y \}$ at
representative states $x$ in Figure 1, will
give the reader a good feel for the geometric
nature of the Bayesian order.

\begin{figure}[h]
\centering\epsfig{figure=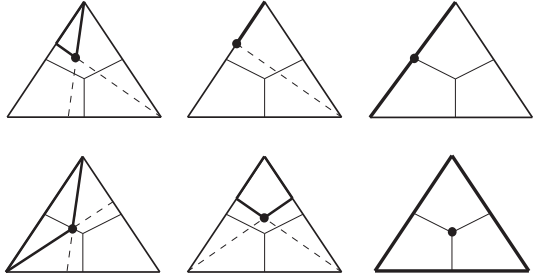,width=200pt}
\caption{Pictures of $\uparrow\!\!x$ for $x\in\Delta^3$.}
\end{figure}

\subsection{The Quantum Case}
The real force of the construction for classical states becomes
apparent in the further development in \cite{Coecke2002}, to show that it can be 
lifted to analogous constructions for \emph{quantum states}. Here,
rather than probability distributions on finite sets, one is looking at 
\emph{mixed states on finite-dimensional Hilbert spaces}. 
Let $\mathcal{H}^n$ denote an $n$-dimensional complex Hilbert
space.
A \em quantum state \em is
a density operator ${\rho:\mathcal{H}^n\rightarrow\mathcal{H}^n}$, i.e.,
a self-adjoint, positive, linear operator with $\mathrm{tr}(\rho)=1.$
The quantum states on $\mathcal{H}^n$ are denoted $\Omega^n$.
A quantum state $\rho$ on $\mathcal{H}^n$
is \em pure \em if
\[\mathrm{spec}(\rho)\subseteq\{0,1\}.\]
The set of pure states is denoted $\Sigma^n$. They
are in bijective correspondence with the one-dimensional subspaces
of $\mathcal{H}^n.$
Classical states are distributions on the
set of pure states $\max(\Delta^n).$
By Gleason's theorem \cite{Gleason}, an analogous
result holds for quantum states: Density operators encode
distributions on $\Sigma^n$.\footnote{Of course, Gleason's theorem also applies to separable infinite-dimensional spaces.}

If our
knowledge about the state of a system
is represented by density operator $\rho$,
then quantum mechanics predicts
the probability that a measurement of observable $e$
yields the value $\lambda\in\mathrm{spec}(e)$. It is
\[\mathrm{pr}(\rho\rightarrow
e_\lambda):=\mathrm{tr}(p_e^\lambda\cdot\rho),\]
where $p_e^\lambda$ is the projection corresponding
to eigenvalue $\lambda$ and $e_\lambda$ is its associated
eigenspace in the \em spectral representation \em of $e$.

Let $e$ be an observable on $\mathcal{H}^n$ with
$\mathrm{spec}(e)=\{1,\ldots,n\}.$ For a quantum state $\rho$ in
$\Omega^n$,
\[\mathrm{spec}(\rho|e):=(\mathrm{pr}(\rho\rightarrow
e_1),\ldots,\mathrm{pr}(\rho\rightarrow
e_n))\in\Delta^n.\]

So what does it mean to say that we have more information
about the system when we have $\sigma\in\Omega^n$ than when we have
$\rho\in\Omega^n$?
It means that there is an observable
$e$ such that (a) the meaurement of $e$ serves as a physical realization of the knowledge
each state imparts to us, and (b) we have a better
chance of predicting the result of the measurement of $e$ in state $\sigma$ than we do in
state $\rho$.
Formally, (a) means
that $\mathrm{spec}(\rho)=\mathrm{Im}(\mathrm{spec}(\rho|e))$ and
$\mathrm{spec}(\sigma)=\mathrm{Im}(\mathrm{spec}(\sigma|e))$ (where the image $\mathrm{Im}$ simply converts a list to the underlying set),
which is equivalent to
requiring $[\rho,e]=0$ and $[\sigma,e]=0$,
where $[a,b]=ab-ba$ is the commutator of operators.

\begin{definition}\em Let $n\geq 2$. For quantum states
$\rho,\sigma\in\Omega^n$, we have $\rho\sqsubseteq \sigma$ iff there
is an observable $e:\mathcal{H}^n\rightarrow\mathcal{H}^n$ such
that $[\rho,e]=[\sigma,e]=0$ and
$\mathrm{spec}(\rho|e)\sqsubseteq\mathrm{spec}(\sigma|e)$ in $\Delta^n$.
\end{definition}
Taking this definition together with our reading of the Bayesian order on classical states, we capture the idea of being able to predict the result of an experiment more confidently on $\sigma$ than on $\rho$ in terms of the less rapid falling off of the values of $\mathrm{spec}(\rho|e)_{i}$ than of $\mathrm{spec}(\sigma|e)_{i}$.

\begin{theorem} $(\Omega^n,\sqsubseteq)$ is a domain with maximal elements
\[\mathrm{max}(\Omega^n)=\Sigma^n\]
and least element $\bot=I/n$, where $I$ is
the identity matrix. Moreover, 
von Neumann entropy
\[ S(\rho) = -\mathrm{tr}(\rho\log\rho)\]
is a measurement of type $\Omega^n\rightarrow[0,\infty)^*.$
\end{theorem}
This order can be characterized in a similar fashion to the Bayesian order on
$\Delta^n$,
in terms
of symmetries and projections. In its symmetric
formulation, \em unitary operators \em on $\mathcal{H}^n$
take the place of permutations on $\{1,\ldots,n\}$,
while the projective formulation of $(\Omega^n,\sqsubseteq)$
shows that each classical projection
$p_i:\Delta^{n+1}\rightharpoonup\Delta^n$
is actually the restriction of a special
quantum projection $\Omega^{n+1}\rightharpoonup\Omega^n$.

\subsection{The Logics of Birkhoff and von Neumann}
Quantum Logic in the sense of Birkhoff and von Neumann  \cite{BvN}
consists of the propositions one can make about
a physical system. Each proposition takes the form
``The value of observable $e$ is contained in
$E\subseteq\mathrm{spec}(e).$'' For
classical systems, the logic
is $\mathcal{P}\{1,\ldots,n\}$,
while for quantum systems it is $\mathbb{L}^n$,
the lattice of (closed) subspaces of $\mathcal{H}^n.$
In each case, implication of propositions is
captured by inclusion, and a fundamental
distinction between
classical and quantum --- that
there are pairs of quantum observables
whose exact values cannot be simultaneously measured at a single
moment in time --- finds lattice theoretic expression:
$\mathcal{P}\{1,\ldots,n\}$ is distributive;
$\mathbb{L}^n$ is not.

The classical
and quantum logics can be \em derived \em
from the Bayesian and spectral orders
using the \em same \em order theoretic construction.

\begin{definition}\em An element $x$ of a dcpo $D$ is \em
irreducible \em when
\[\bigwedge(\uparrow\!\!x\cap\max(D))=x\]
The set of irreducible elements in $D$ is written $\mbox{Ir}(D).$
\end{definition}

The order dual of a poset $(D,\sqsubseteq_D)$ is written $D^*$; its
order is $x\sqsubseteq y\Leftrightarrow y\sqsubseteq_D x.$

The following result is proved in \cite{Coecke2003}.

\begin{theorem} For $n\geq 2$, the classical lattices arise as
\[\mathrm{Ir}(\Delta^n)^*\simeq\mathcal{P}\{1,\ldots,n\}\setminus\{\emptyset\},\]
and the quantum lattices arise as
\[\mathrm{Ir}(\Omega^n)^*\simeq\mathbb{L}^n\setminus\{0\}.\]
\end{theorem}

\begin{figure}[h]
\centering\epsfig{figure=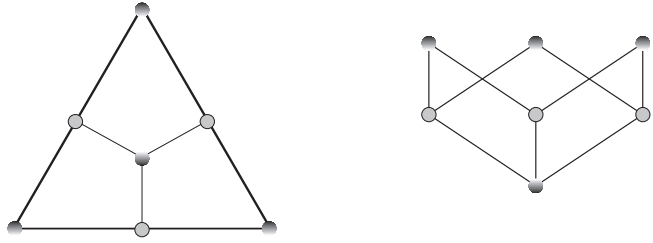,width=250pt}
\caption{The irreducibles of $\Delta^3$ with the
corresponding Hasse diagram.}
\end{figure}

\subsection{Discussion}
The foregoing development has been quite technical, but the  underlying programme which these ideas illustrate has a clear conceptual interest. The broad agenda of developing a unified quantitative/qualitative theory of information, applicable to a wide range of situations in logic and computation, is highly attractive, and likely to lead to new perspectives on information in general.

Our discussion thus far has largely been couched in terms of \emph{static} theories, although we have already hinted at the importance of agents and explicit dynamics. We now turn to \emph{interactive models} of logic and computation.

\section{Games, Logical Equilibria and Conservation of Information Flow}

In this Section and the next, we shall discuss some dynamical theories of computation which are explicitly based on \emph{interaction} between agents, and which expose a structure of information flow which is both \emph{geometrical} and \emph{logical} in character. These theories, which go under the names of \emph{Game Semantics} and \emph{Geometry of Interaction}, have played a considerable r\^ole in recent work on the semantics both of programming languages, and of logical proofs.

\subsection{Changing Views of  Computation}
To set the scene, we begin by recalling how perspectives on computation have changed since the first computers appeared.
The early practice of computing can be pictured as in Figure 3.
\begin{figure}
\centering\epsfig{figure=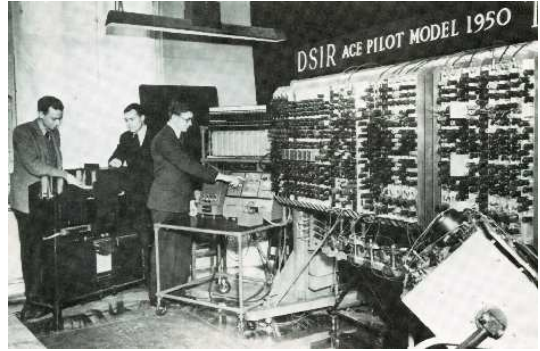,width=200pt}
\caption{Computing ``in the isolation ward''.}
\end{figure}
This is the era of stand-alone machines and programs: computers are served by an elite priesthood, and have only a narrow input-output interface with the rest of the world.

\paragraph{First-generation models of computation}
Given this limited vision of computing, there is a very natural abstraction of computation, in which 
programs are seen as computing \emph{functions} or \emph{relations} 
from inputs to outputs.\footnote{This is the exactly the point of view on which, as we have seen, program logics such as Hoare Logic and Dynamic Logic are based.}
\begin{center}
\psset{unit=1in,cornersize=absolute,dimen=middle}%
\begin{pspicture}(-1.142857,-0.571429)(2.857143,0.571429)%
% dpic version 08.Jul.05 for PSTricks 0.93a
\psframe[fillstyle=solid,fillcolor=yellow,linecolor=blue](0,-0.571429)(1.714286,0.571429)
\rput(0.857143,0){Computation}
\psline[arrowsize=0.05in 0,arrowlength=2,arrowinset=0]{->}(-1.142857,0)(0,0)
\uput{0.5ex}[u](-0.571429,0){Input Data}
\psline[arrowsize=0.05in 0,arrowlength=2,arrowinset=0]{->}(1.714286,0)(2.857143,0)
\uput{0.5ex}[u](2.285714,0){Output}
\end{pspicture}%

\end{center}
These models live on the existing intellectual inheritance from discrete mathematics and logic.
\emph{Time} and \emph{processes} lurk in the background, but are largely suppressed.

\paragraph{Computation in the Age of the Internet}
As we know, the technology has changed dramatically. Even a conventional Distributed Systems picture, as illustrated in Figure~4, which has been common-place for the last 20 years, tells a very different story.
\begin{figure}
\begin{center}
\resizebox{15cm}{10cm}{\includegraphics{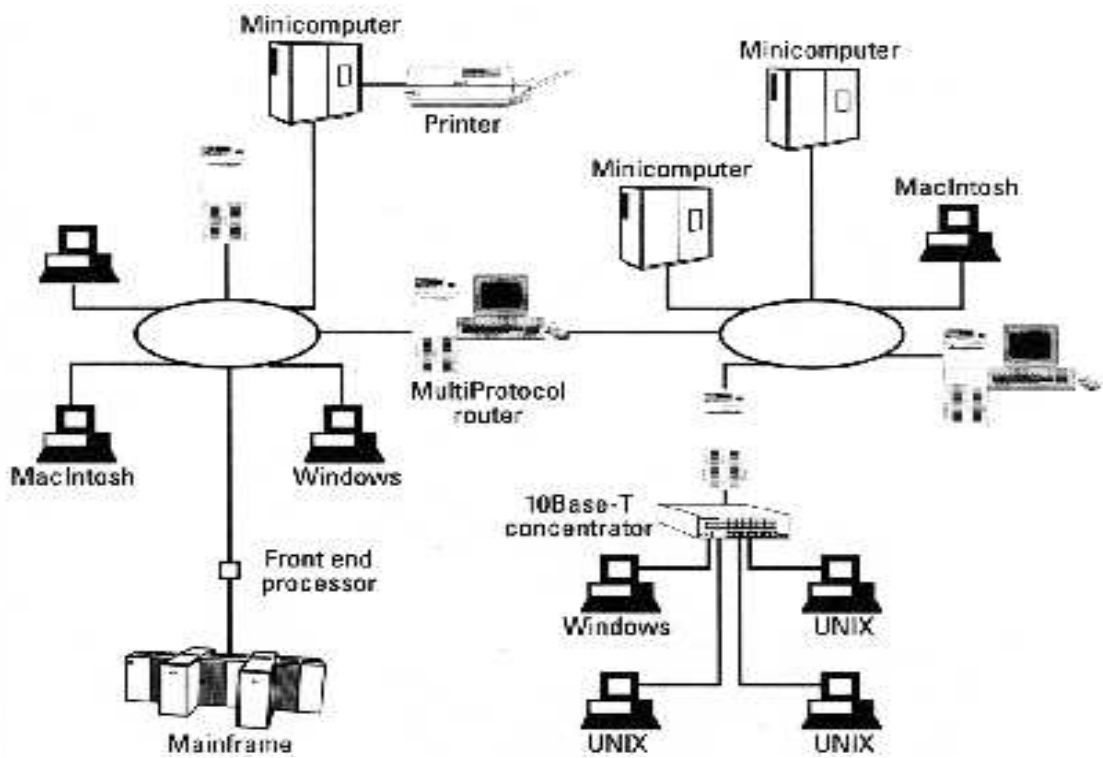}}
\end{center}
\caption{Distributed Computing}
\end{figure}
We have witnessed the progression
\begin{center}
multitasking $\rightarrow$ distributed systems $\rightarrow$ Internet
$\rightarrow$ ``mobile'' and ``global'' computing
\end{center}
Key features of this unfolding new computational universe include:
\emph{agents interacting} with each other, and 
\emph{information flowing} around the system.

The insufficiency of the first-generation models of computation for this new computational environment is evident.
The old concepts fail to match the modern world of computing and its concerns:
\begin{description}
\item[Robustness] in the presence of failures.
\item[Atomicity] of transactions.
\item[Security] of information flows. 
\item[Quality] of user interface.
\item[Quantitative] aspects.
\end{description}

\paragraph{Processes vs. Products}
We see a shift in emphasis and importance between \emph{How} we compute \textit{vs.}~\emph{What} we compute.
Processes were in the background, but now come to the fore: the ``how''
\emph{becomes} the new ``what''.

This leads ineluctably to the need for \textbf{Second-generation models} of computation, and in particular
\emph{Process Models} such as Petri nets, Process Algebra, etc.
Whereas 1st-generation models lived off the intellectual inheritance from
mathematics and logic, there is no adequate pre-existing theory of 
\emph{processes} or \emph{agents}, \emph{interaction}, and \emph{information flow}, as we see by considering the following questions (which have already been mentioned in Section~1):
\begin{itemize}
\item \emph{What} is computed? 
\item What \emph{is} a process? 
\item What are the analogues to
Turing-completeness, universality?
\end{itemize}
There are indeed a plethora of models, but no definitive conceptual analysis, comparable
to Turing's analysis of computation in its ``classical'' sense:
not least, perhaps, because it is indeed \emph{a harder problem}!

\subsection{Some New Perspectives}
Instead of isolated systems, with rudimentary interactions with their
environment, the standard unit of description or design becomes a
\emph{process} or \emph{agent}, the essence of whose behaviour is
\emph{how it interacts} with its environment.
\begin{center}
\resizebox{3cm}{3cm}{\includegraphics{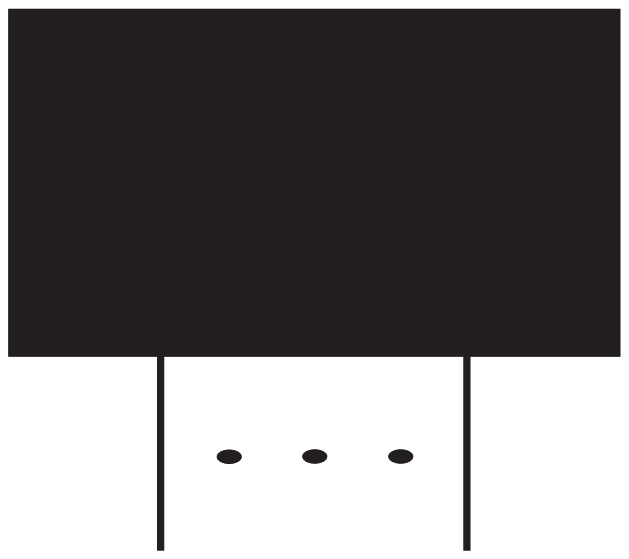}}
\end{center}
Who is the System? Who is the Environment? This depends on  point of view. We may designate some agent or group of agents as the System currently under consideration, with everything else as the Environment; but it is always possible to contemplate a r\^{o}le interchange, in which the Environment becomes the System and vice versa. (This is, of course, one of the great devices, and imaginative functions, of creative literature).
This \emph{symmetry} between System and Environment carries a first clue that there is some structure here;  it will lead us to a key \emph{duality}, and a deep
connection to logic.

\subsection{Interaction}

Complex behaviour arises as the global effect of a \emph{system}
of \emph{interacting agents} (or processes).

The key building block is the agent.
The key operation is \emph{interaction} -- plugging agents
together so that they interact with each other
\begin{center}
\resizebox{5cm}{2.5cm}{\includegraphics{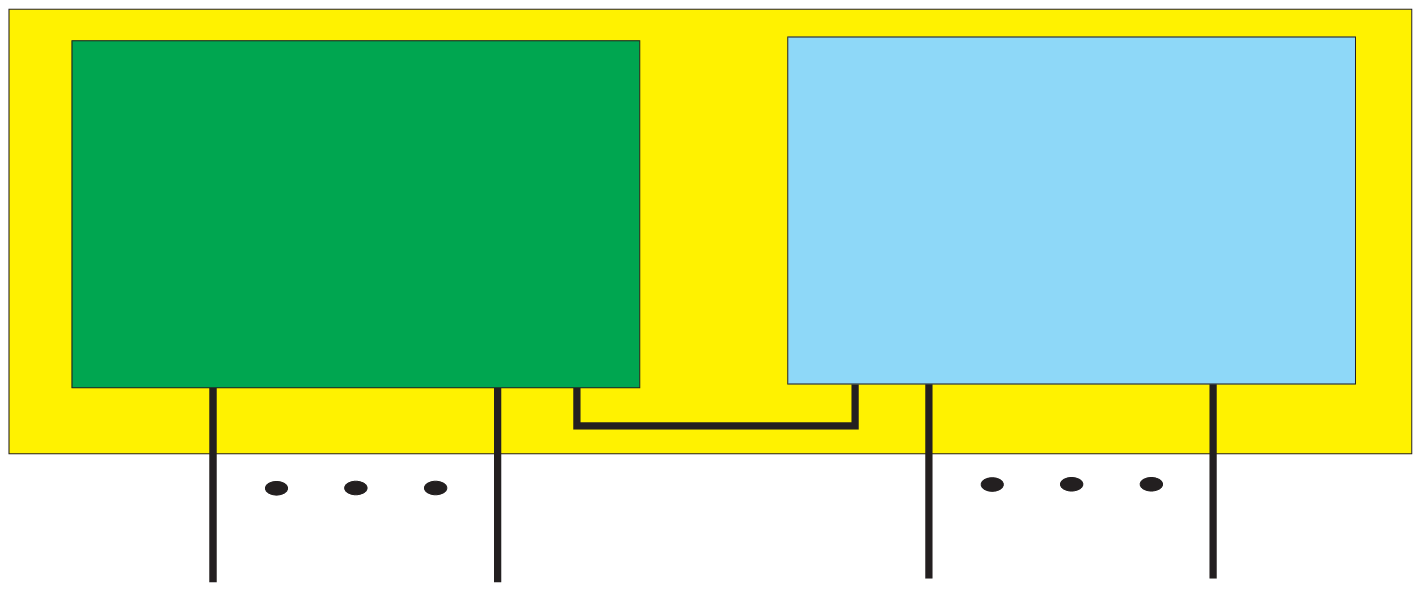}}
\end{center}

This conceptual model works at all ``scales'' :
\begin{itemize}
\item Macro-scale: processes in operating systems, software agents on the
Internet, transactions. 
\item Micro-scale: how programs are implemented (subroutine call-return
protocols, register transfer) all the way down into hardware. 
\end{itemize}
It is applicable both to \emph{design} (synthesis) and to \emph{description} (analysis);
to \emph{artificial} and to \emph{natural} information-processing systems. 

There are of course large issues lurking here, e.g. in the realm of ``Complex Systems'': 
\emph{emergent behaviour} and even \emph{intelligence}. 
Is is helpful, or even feasible, to understand this complexity \emph{compositionally}?
We need new conceptual tools, new theories, to help us analyze and synthesize 
these systems,
to help us to \emph{understand} and to \emph{build}.

\subsection{Towards a ``Logic of Interaction''}
Specifying and reasoning about the behaviour of computer programs
takes us into the realm of logic.
For the first-generation models, logic could be taken ``as it was''---static
and timeless.
For our second-generation models, getting an adequate account---a genuine
``logic of interaction''---may require a fundamental reconceptualization
of logic itself.
This radical revision of our view of logic is happening anyway---prompted
partly by the applications, and partly by ideas arising within  logic.

\subsubsection{The Static Conception of Logic}

We provide an unfair caricature of the standard logical idea of tautology to make our point.
The usual ``static'' notion of tautology is as  ``a statement which is vacuously true because it is compatible with all states of affairs''.
\begin{center}
$A \vee \neg A$ 
\end{center}
``It is raining \emph{or} it is not raining''---truth-functional 
semantics. 
This is illustrated (subversively) in Figure 5.
\begin{figure}[h]
\begin{center}
\begin{tabular}{cccccc}
\resizebox{4.5cm}{3.5cm}{\includegraphics{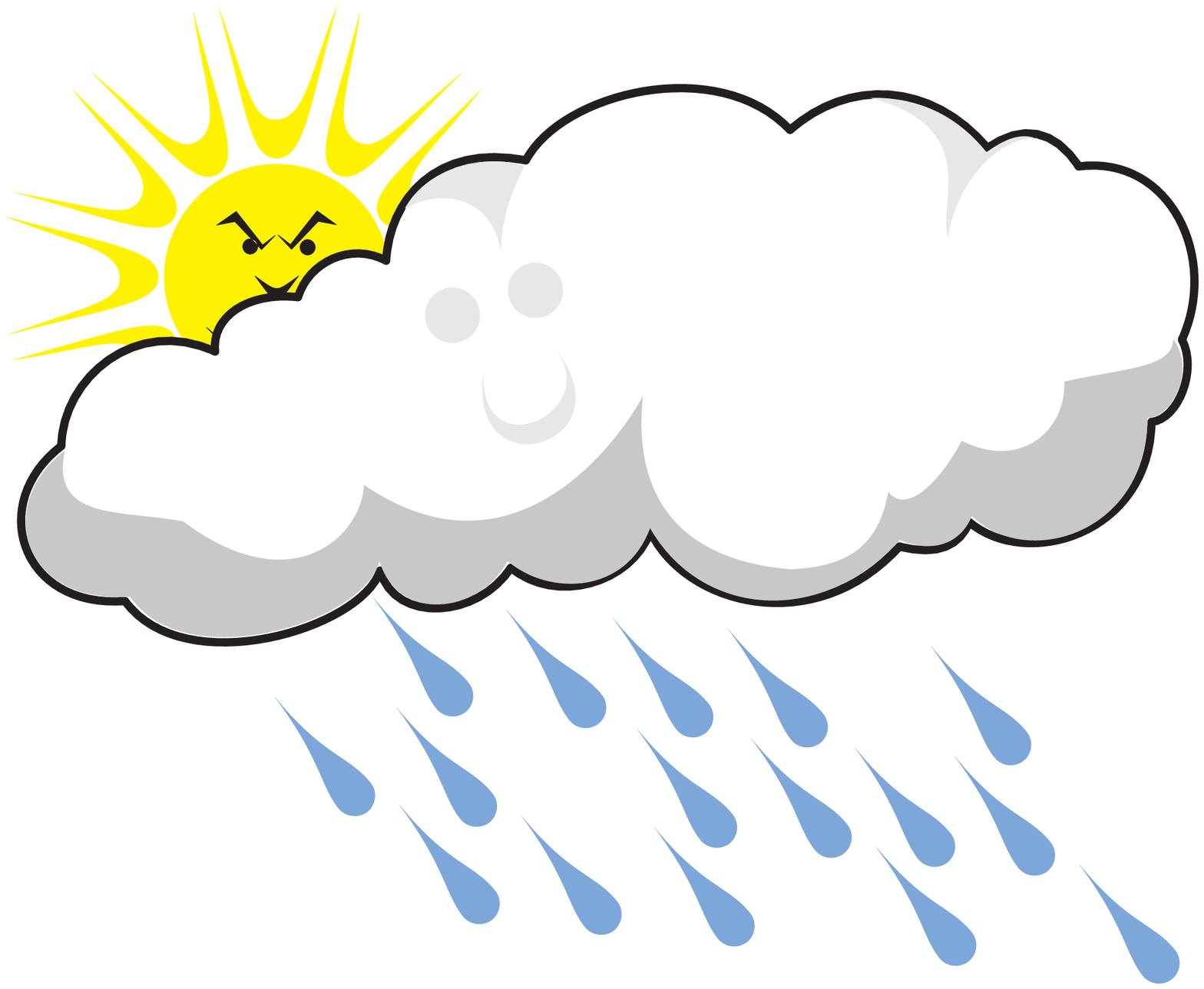}}&\ \ &
\resizebox{4.5cm}{3.5cm}{\includegraphics{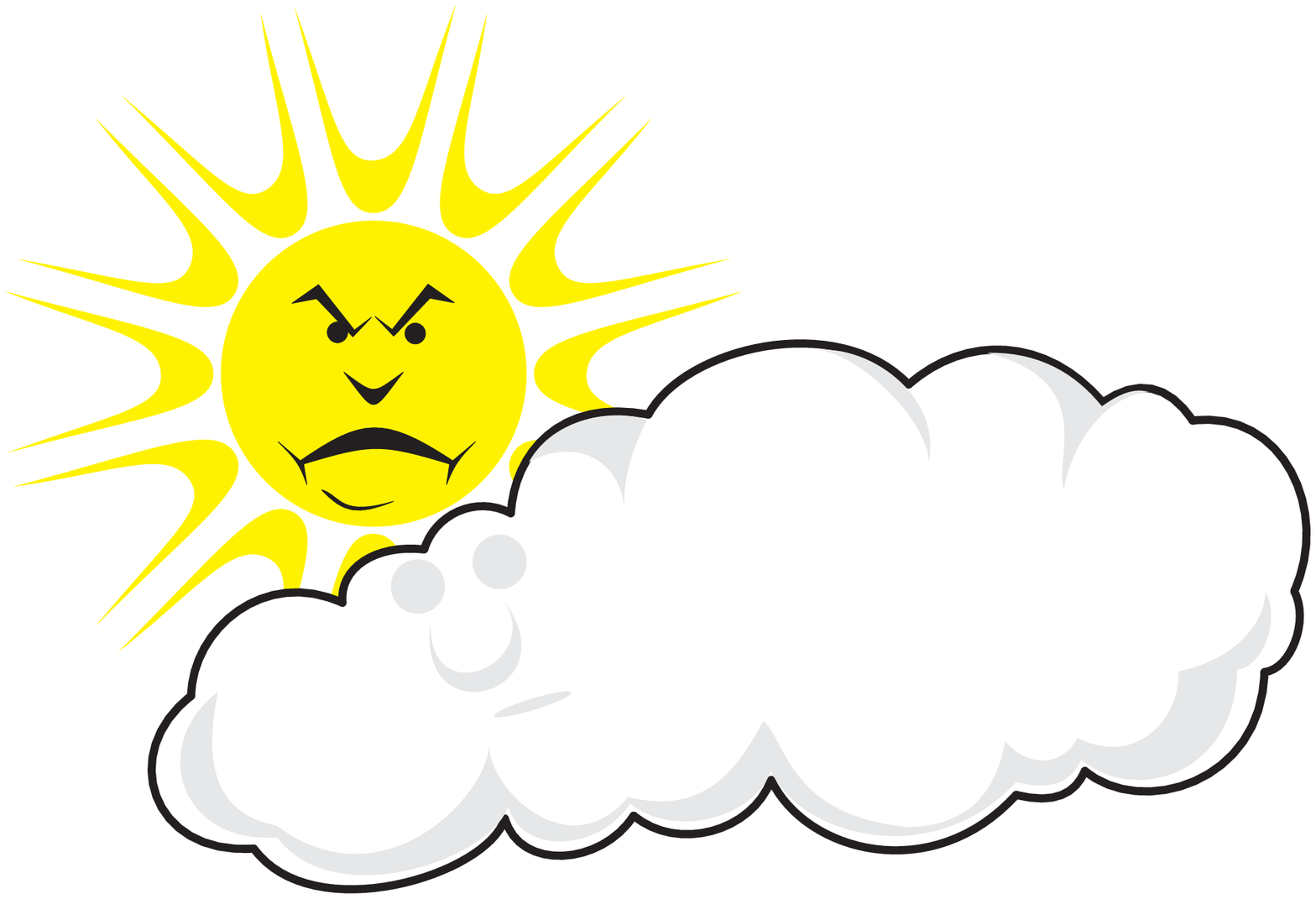}}
&\ \  &
\resizebox{4.5cm}{3.5cm}{\raisebox{1em}{\includegraphics{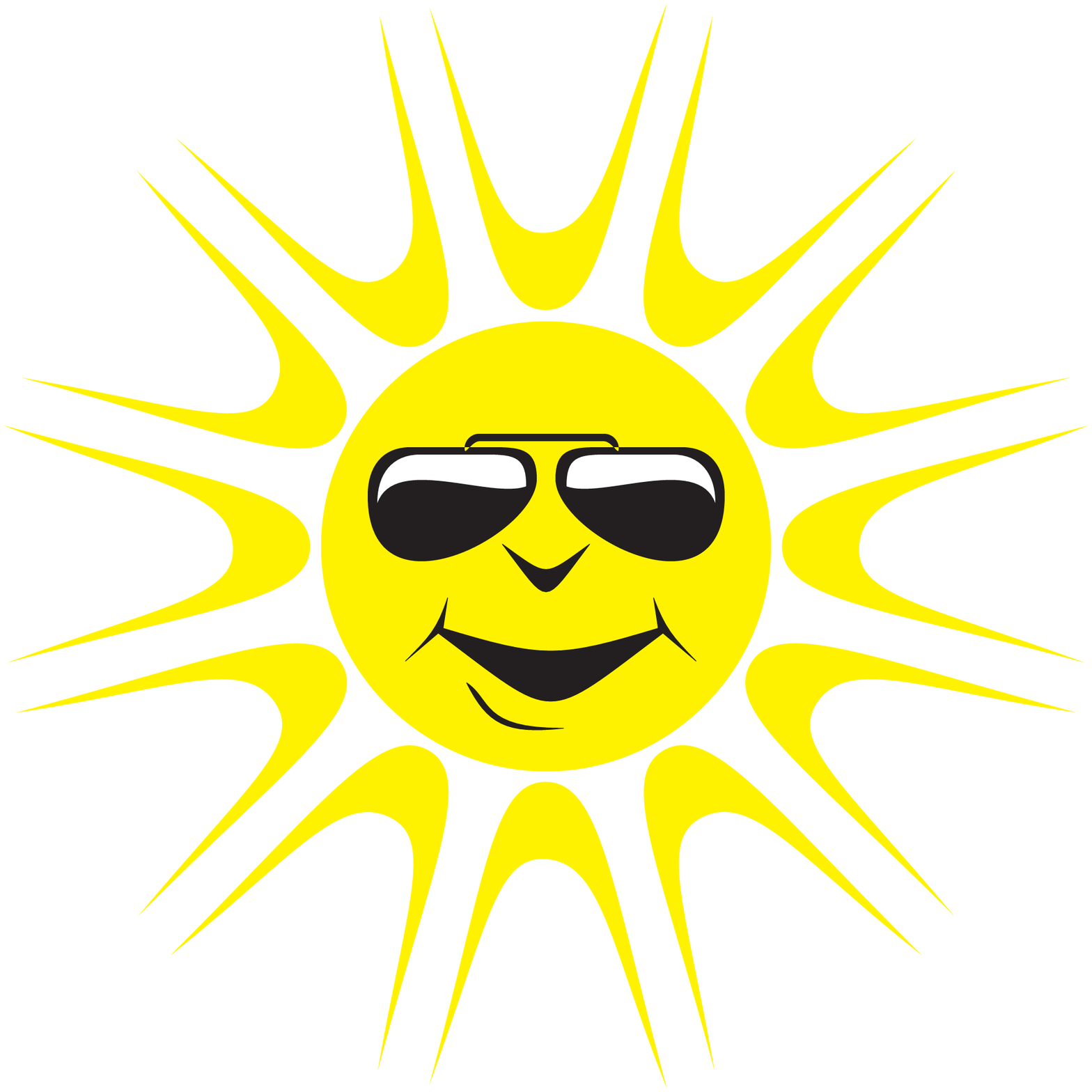}}}
&\ \ \ \ \ 
\end{tabular}
\end{center}
\caption{Tertium non datur?}
\end{figure}
But what could a \emph{dynamic notion of tautology} look like?

\subsubsection{The Copy-Cat Strategy}
We begin with a little fable, illustrated by Figure 6:
\begin{center}
\fbox{How to beat an International Chess Grandmaster by the power of pure logic}
\end{center}
\begin{figure}
\begin{center}
\epsfig{file=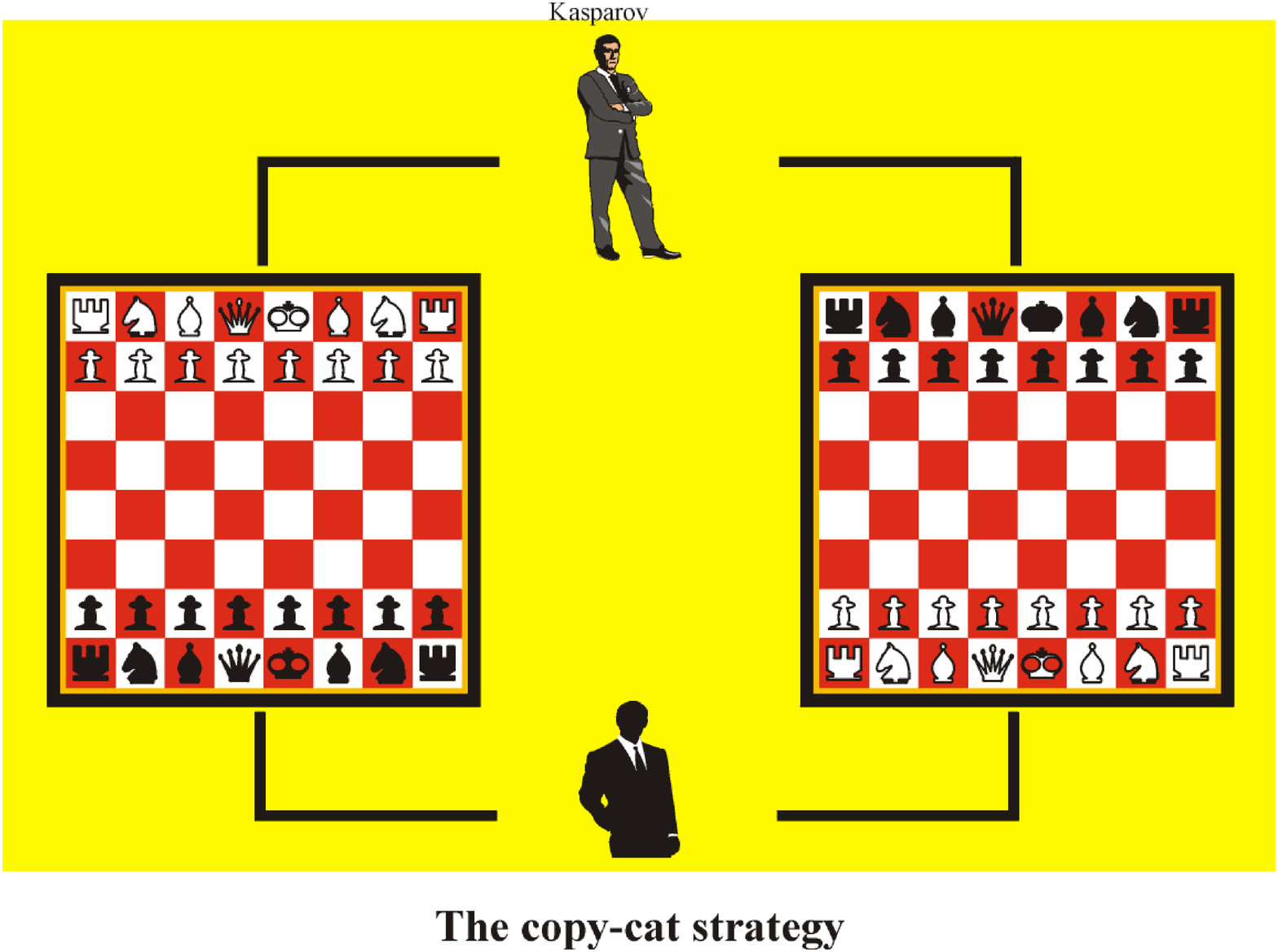,height=3in,width=4.5in}
\end{center}
\caption{How to beat a Grandmaster}
\end{figure}
Since we are relying on logic, rather than on any talent at Chess, we proceed as follows. We arrange to play two games of Chess with the grandmaster, say Gary Kasparov, once as White and once as Black. Moreover, we so arrange matters that we start with the game in which we play as Black. Kasparov makes his opening move; we respond by playing the \emph{same} move in the \emph{other} game---this makes sense, since we are playing as White there. Now Kasparov responds (as Black) to our move in that game; and we copy that response back in the first game. We simply proceed in this fashion, copying the moves that our opponent makes in one board to the other board. The net effect is that \emph{we play the same game twice---once as White, and once as Black}. (We have essentially made Kasparov play against himself). Thus, whoever wins that game, we can claim a win in one of our games against Kasparov! (Even if the game results in a stalemate, we have done as well as Kasparov over the two games---surely still a good result!)\footnote{Our fable is actually recorded as having happened at least once in the chronicles of Chess. Two players conspired to play this copy-cat strategy against Alekhine in the 1920's. Alekhine realized what was happening, and made a tempting offer of a sacrifice to one of his opponents. That opponent was not able to resist such a coup against the great Alekhine, and departed from the copy-cat strategy to swallow the bait. Then the symmetry was broken, and Alekhine was able to win easily in both games. Thus we are reminded of the familiar truth, that logic rarely prevails over psychology in ``real life''.}

Of course, this idea has nothing particularly to do with Chess. It can be applied to any two-person game of a very general form. We shall continue to use Chess-boards to illustrate our discussion, but this underlying generality should be kept in mind.

What are the salient features which can be extracted from this example?
\begin{description}
\item[A dynamic tautology] 
There is a sense (which will shortly be made more precise) in which the copy-cat strategy can be seen as a \emph{dynamic version} of the tautology $A \vee \neg A$. Note, indeed, that an essential condition for being able to play the copy-cat is that the r\^oles of the two players are inter-changed on one board as compared to the other. Note also the disjunctive quality of the argument that we must win in one or other of the two games. But the copy-cat strategy is a \emph{dynamic process}: a two-way channel which maintains the correlation between the plays in the two games.
\item[Conservation of information flow]
The copy-cat strategy does not \emph{create} any information; it reacts to the environment in such a way that information is conserved. It ensures that exactly the same information flows out to the environment as flows in from it. Thus one gets a sense of logic appearing in the form of \emph{conservation laws for information dynamics}.
\item[The power of copying]
Another theme which appears here, and which we will see more of later, concerns the surprising power of simple processes of copying information from one place to another. Indeed, as we shall eventually see, such processes are \emph{computationally universal}.
\item[The geometry of information flow]
From a dynamical point of view, the copy-cat strategy realizes a channel between the two game boards, by performing the \emph{actions} of copying moves. But there is also some implicit \emph{geometry} here. Indeed, the very idea of two boards laid out side by side appeals to some basic underlying spatial structure. In these terms, the copy-cat channel can also be understood geometrically, as creating a graphical link between these two spatial locations.These two points of view are complementary, and link the logical perspective to powerful ideas arising in modern geometry and mathematical physics.
\end{description}
To provide further evidence that the copy-cat strategy embodies more substantial ideas than might at first be apparent, we consider varying the scenario. Consider now the case where we play against Kasparov on \emph{three boards}; one as Black, two as White.
$$\xymatrix{
  *\txt{Kasparov} & & *\txt{Kasparov} & *\txt{Kasparov} \\
  *\txt{{
    \begin{tabular}{|c|}
      \hline\\
      \mbox{\hspace{2em}B\hspace{2em}}\\
      \hline\\
      W\\
      \hline
    \end{tabular}}}& & 
  *\txt{{
    \begin{tabular}{|c|}
      \hline\\
      \mbox{\hspace{2em}W\hspace{2em}}\\
      \hline\\
      B\\
      \hline
    \end{tabular}}} & 
  *\txt{{
    \begin{tabular}{|c|}
      \hline\\
      \mbox{\hspace{2em}W\hspace{2em}}\\
      \hline\\
      B\\
      \hline
    \end{tabular}}} \\
 & {\cdot}\ar@{-}[ul]+D\ar@{-}[ur]+D\ar@{-}[urr]+D
}$$
Does the Copy-Cat strategy still work here? In fact, we can easily see that it does \emph{not}. Suppose Kasparov makes an opening move $m_{1}$ in the left-hand board where he plays as White; we copy it to the board where we play as White; he responds with $m_{2}$; and we copy $m_{2}$ back to the board where Kasparov opened. So far, all has proceeded as in our original scenario. But now Kasparov has the option of playing a \emph{different} opening move, $m_{3}$ say, in the rightmost board. We have no idea how to respond to this move; nor can we copy it anywhere, since the board where we play as White is already ``in use''. This shows that these simple ideas already lead us naturally to the setting of a \emph{resource-sensitive} logic, in which in particular the Contraction Rule, which can be expressed as $A \rarr A \wedge A$ (or equivalently as $\neg A \vee (A \wedge A)$) cannot be assumed to be valid.

What about the other obvious variation, where we play on two boards as White, and one as Black?
$$\xymatrix{
  *\txt{Kasparov} & *\txt{Kasparov} &  & *\txt{Kasparov} \\
  *\txt{{
    \begin{tabular}{|c|}
      \hline\\
      \mbox{\hspace{2em}B\hspace{2em}}\\
      \hline\\
      W\\
      \hline
    \end{tabular}}}& 
  *\txt{{
    \begin{tabular}{|c|}
      \hline\\
      \mbox{\hspace{2em}B\hspace{2em}}\\
      \hline\\
      W\\
      \hline
    \end{tabular}}} & &
  *\txt{{
    \begin{tabular}{|c|}
      \hline\\
      \mbox{\hspace{2em}W\hspace{2em}}\\
      \hline\\
      B\\
      \hline
    \end{tabular}}} \\
 & & {\cdot}\ar@{-}[ull]+D\ar@{-}[ul]+D\ar@{-}[ur]+D
}$$
It seems that the copy-cat strategy \emph{does} still work here, since we can simply ignore one of the boards where we play as White. However, a geometrical property of the original copy-cat strategy has been lost, namely a \emph{connectedness} property, that information flows to every part of the system. This at least calls the corresponding logical principle of Weakening, which can be expressed as $A \wedge A \rarr A$, (or equivalently as $\neg A \vee \neg A \vee A$) into question.

We see from these remarks that we are close to the realm of Linear Logic and its variants; and, mathematically, to the world of monoidal (rather than cartesian) categories.

\subsubsection{Game Semantics}
These ideas find formal expression in \emph{Game Semantics}.
Games play the role of:
\begin{itemize}
\item Interface types for computation modules 
\item Propositions with dynamic content.
\end{itemize}
In particular,
2-person games capture the duality of:
\begin{itemize}
\item Player \textit{vs.} Opponent 
\item System \textit{vs.} Environment.
\end{itemize}

\paragraph{Agents are strategies}
In this setting, we model our agents or processes as \emph{strategies}
for playing the game. 
These strategies \emph{interact} by playing against each other. 
We obtain a notion of correctness which is \emph{logical} in character
in terms of the idea of \emph{winning} strategy---one which is guaranteed
to reach a successful outcome however the environment behaves. 
This in a sense replaces (or better, \emph{refines}) the logical notion
of ``truth'': winning strategies are our dynamic version of tautologies
(more accurately, of \emph{proofs}).

\paragraph{Building complex systems by combining games}
We shall now see how games can be combined to produce more complex
behaviours while retaining control over the interface. 
This provides a basis for the \emph{compositional} understanding
of our systems of interacting agents---understanding the behaviour
of a complex system in terms of the behaviour of its parts. 
This is crucial for both analysis and synthesis, \textit{i.e.}
for both description
and design. 
These operations for building games can be seen as (dynamic forms of)
``type constructors'' or ``logical connectives''. 
(The underlying logic here will in fact be  Linear Logic).

\paragraph{Duality---``Linear Negation''}
$A^{\perp}$ --- interchange r\^{o}les of Player and Opponent (reflecting the
symmetry of interaction).
\begin{center}
\resizebox{4in}{2in}{\includegraphics{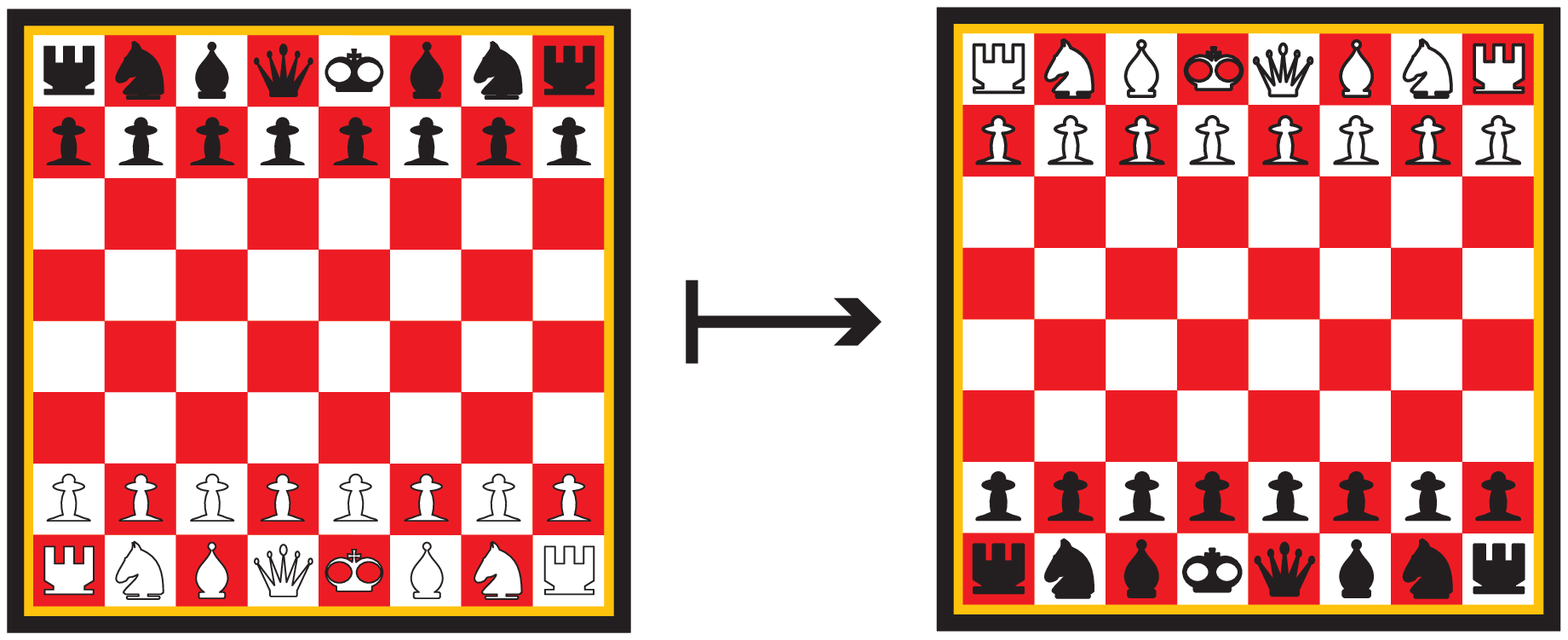}}
\end{center}
Note that, with this interpretation, negation is involutive:
\[ A^{\perp \perp} = A . \]

\paragraph{Tensor --- ``Linear conjunction''}
\begin{center}
\resizebox{3in}{2in}{\includegraphics{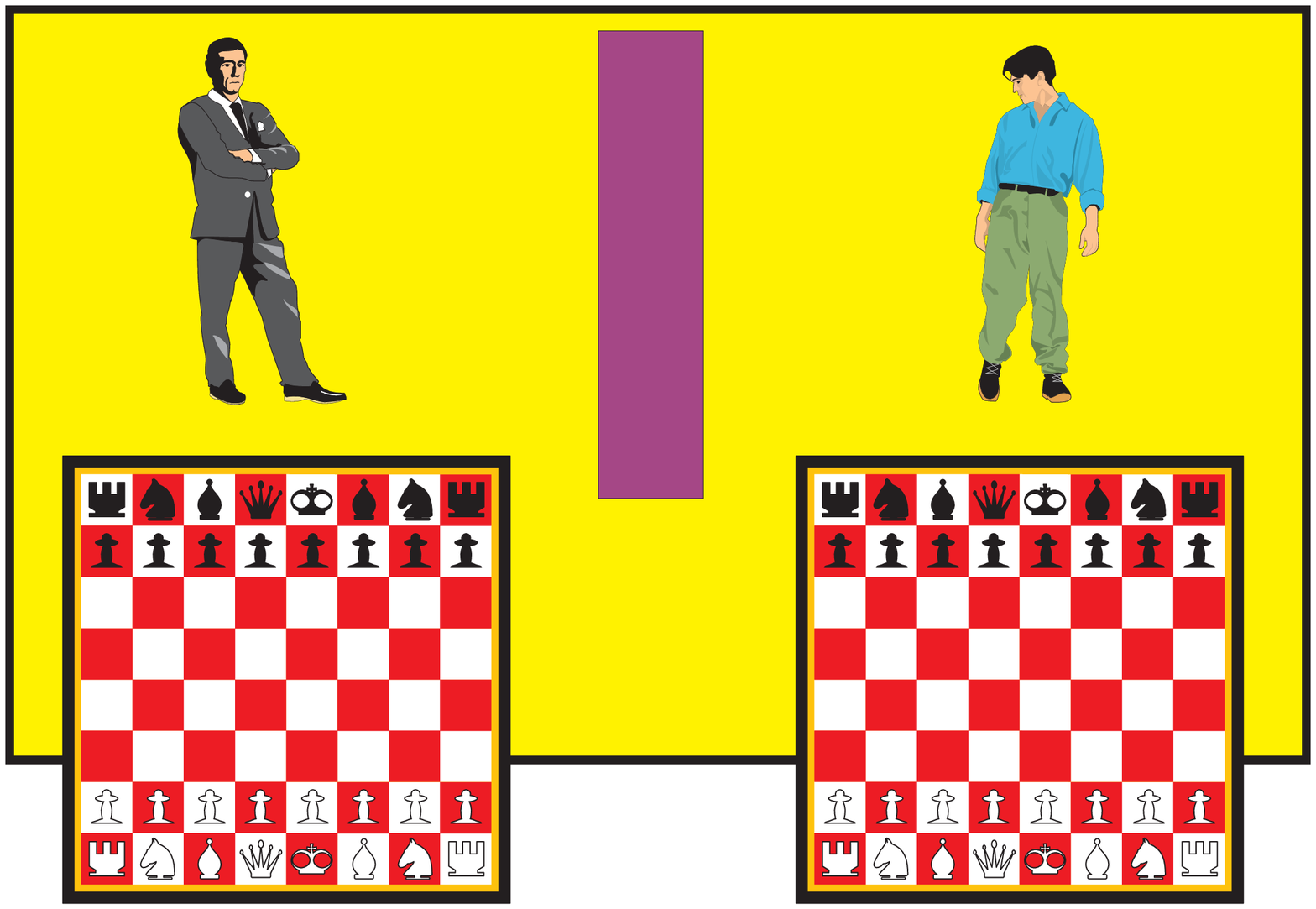}}\\
$A \otimes B$
\end{center}
The idea here is that we combine the two game boards into one system, \emph{without any information flow between the two sub-systems}. (This is the significance of the ``wall'' separating our two players, who we shall refer to as Gary (Kasparov) and Nigel (Short)). This connective has a conjunctive quality, since we must independently be able to play (and to win) in each conjunct. Note however, that there is no constraint on information flow for the environment, as it plays against this compound system. 

\paragraph{Par --- ``Linear disjunction''}
\begin{center}
\resizebox{3in}{2in}{\includegraphics{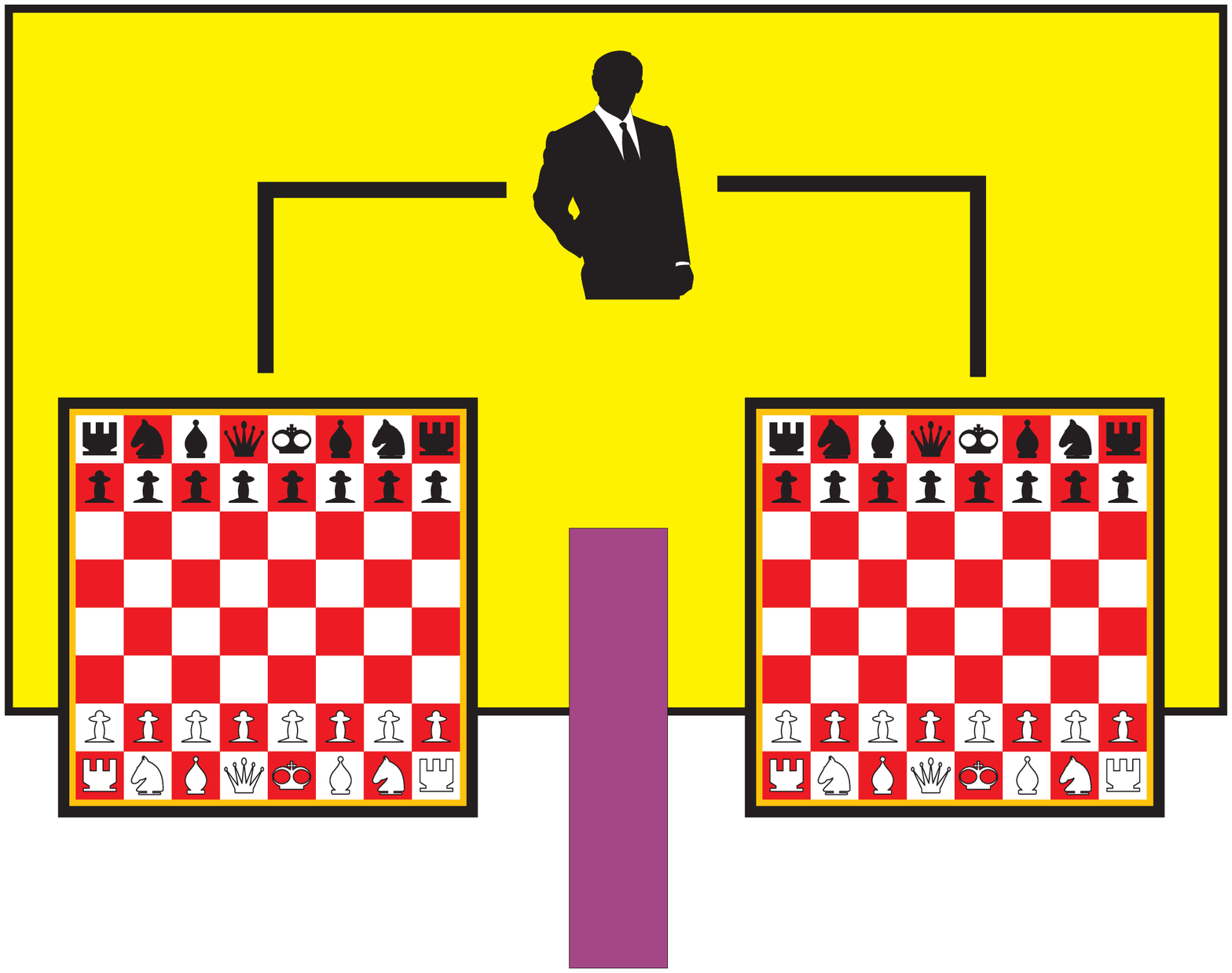}}\\
$A \llpar B$
\end{center}
In this case, we have two boards, but one player (who we refer to as the Copy-Cat), indicating that we \emph{do} allow information flow for this player between the two game boards. This for example allows information revealed in one game board by the Opponent to be used against him on the other game board---as exemplified by the copy-cat strategy. However, note that the wall appears on the environment's side now, indicating that the environment is constrained to play separately on the two boards, with no communication between them.

Thus we have a De Morgan duality between these two connectives, mediated by the Linear negation:
\[ \begin{array}{rcl}
(A \otimes B)^{\perp} & = &  A^{\perp} \llpar B^{\perp} \\
(A \llpar B)^{\perp} & = &  A^{\perp} \otimes B^{\perp} 
\end{array} \]
The idea is that on one side of the mirror of duality (Player/System for the Tensor, Opponent/Environment for the Par), we have the constraint of no information flow, while on the other side, we do have information flow.

We can now  reconstrue the Copy-Cat strategy in logical terms:
\begin{center}
\resizebox{3in}{2in}{\includegraphics{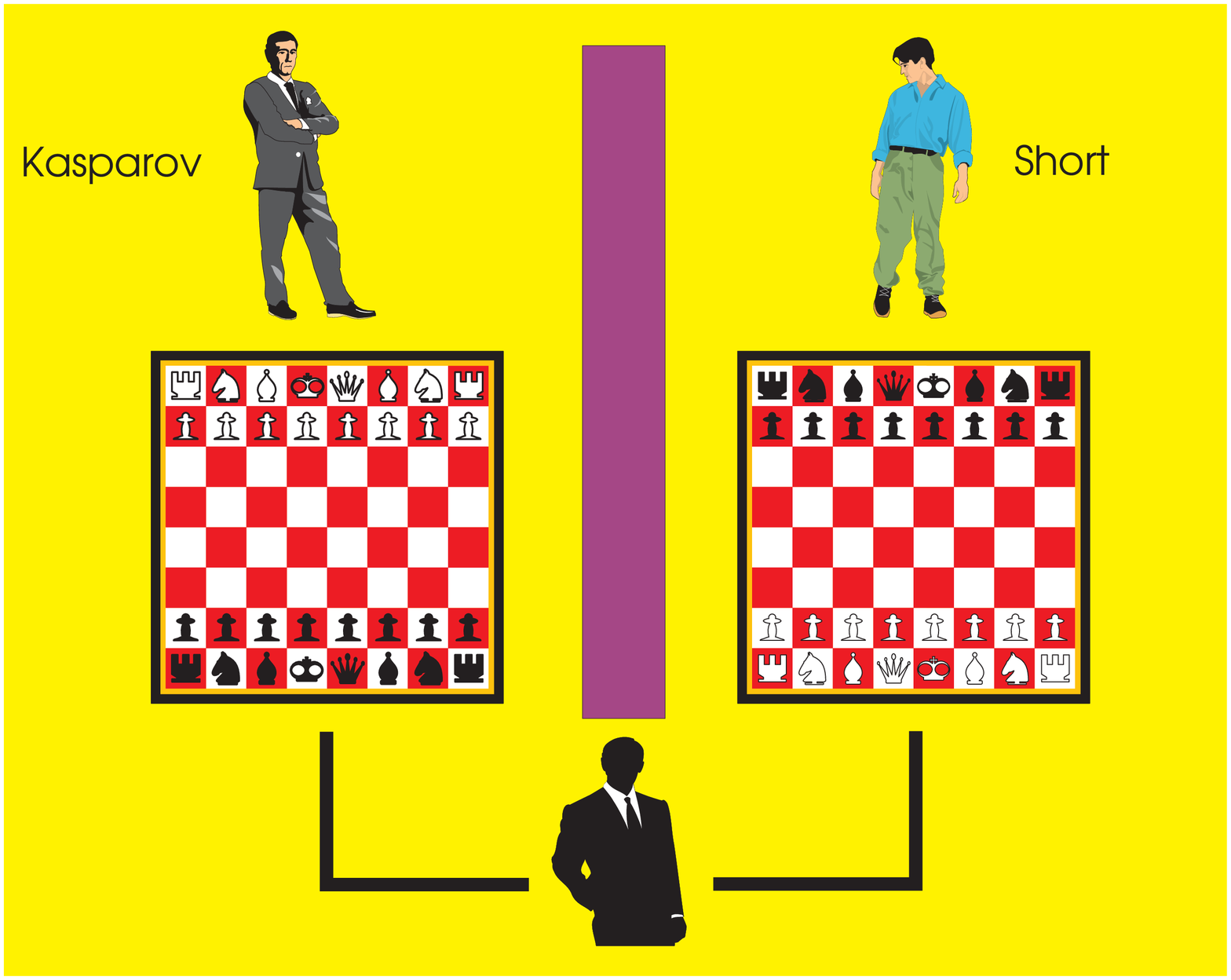}}
\end{center}
We see that it is indeed a winning strategy for 
$A^{\perp} \llpar A$. 
Moreover, we can define $A \llto B$ (``Linear implication'') by
\[ A \llto B  \; \equiv \; A^{\perp} \llpar B \]
(\textit{cf.} $A \supset B  \equiv  \neg A \vee B$.)
Then the copy-cat strategy becomes the canonical proof of the most basic tautology of all: $A \llto A$.

The information flow possibilities of Par receive a more familiar logical interpretation in terms of the Linear implication; namely, that we can use information about the antecedent in proving the consequent (and conversely with respect to their negations, if we think of proof by contraposition).

Thus an entire ``linearized'' logical structure opens up before us, with a natural interpretation in terms of the dynamics of information flow.

\subsubsection{Interaction}
We now turn to a key step in the development: the modelling of \emph{interaction} itself.
Constructors create ``potentials'' for interaction; the operation of plugging
modules together so that they can communicate with each other \emph{releases} this potential into
\emph{actual computation}.
\begin{center}
\resizebox{4.5in}{1.5in}{\includegraphics{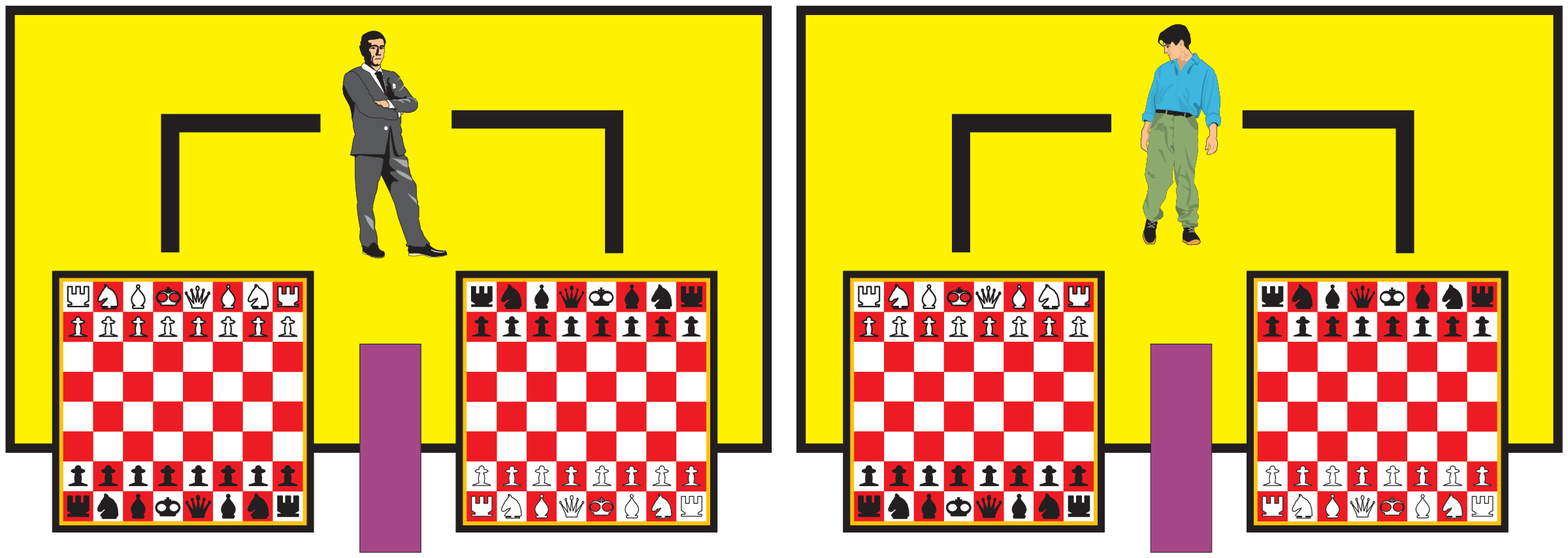}}\\
$A^{\perp} \llpar B \; \equiv \; A \llto B \qquad \qquad \qquad \qquad
B^{\perp} \llpar C \; \equiv \; B \llto C$
\end{center}
Here we see two separate sub-systems, each with a compound structure, expressed by the \emph{logical types of their interfaces}. What these types tell us is that these systems are \emph{composable}; in particular, the \emph{output type} of the first system, namely $B$, matches the input type of the second system. Note that this ``logical plug-compatibility'' makes essential use of the duality, just as the copy-cat strategy did. What makes Gary (the player for the first system) a fit partner for interaction with Nigel (the player for the second system), is that they have \emph{complementary views} of their locus of interaction, namely $B$. Gary will play in this type ``positively'', as Player (he sees it as $B$), while Nigel will play ``negatively'', as Opponent (he sees it as $B^{\perp}$). Thus each will become part of the environment of the other---part of the potential environment of each will be realized by the other, and hence part of the \emph{potential}  behaviour of each will become \emph{actual} interaction.

This leads to a dynamical interpretation of the fundamental operation of \emph{composition}, in mathematical terms:
\[
\begin{diagram}[loose,height=.8em,width=3em]
& A & \rTo^{{\rm Gary}} & B & \rTo^{{\rm Nigel}} & C \\
 \ & & & \hLine & & & \ \\
& A & & \rTo^{{\rm Gary} ; {\rm Nigel}} & & C
\end{diagram} \]
or of the \emph{Cut rule}, in logical terms:
\[ \mbox{Cut:} \qquad \frac{\vdash \Gamma , A \;\; \vdash \llnot{A}, \Delta}{\Gamma , \Delta} \]

\begin{center}
\setlength{\unitlength}{0.0500em}%
\resizebox{2.4in}{1.2in}{\begin{picture}(520,295)(80,365)
\thicklines
\put(540,520){\line( 0,-1){140}}
\put(480,520){\line( 0,-1){140}}
\put(200,520){\line( 0,-1){140}}
\put(140,520){\line( 0,-1){140}}
\put(260,520){\line( 0,-1){ 40}}
\put(260,480){\line( 1, 0){160}}
\put(420,480){\line( 0, 1){ 40}}
\put(400,520){\framebox(160,100){}}
\put(120,520){\framebox(160,100){}}
\put( 80,460){\framebox(520,200){}}
\put(495,435){\makebox(0,0)[lb]{\raisebox{0pt}[0pt][0pt]{ $\ldots$}}}
\put(500,365){\makebox(0,0)[lb]{\raisebox{0pt}[0pt][0pt]{ $\Delta$}}}
\put(160,365){\makebox(0,0)[lb]{\raisebox{0pt}[0pt][0pt]{ $\Gamma$}}}
\put(155,435){\makebox(0,0)[lb]{\raisebox{0pt}[0pt][0pt]{ $\ldots$}}}
\put(423,490){\makebox(0,0)[lb]{\raisebox{0pt}[0pt][0pt]{ $A^{\bot}$}}}
\put(212,490){\makebox(0,0)[lb]{\raisebox{0pt}[0pt][0pt]{ $A$}}}
\put(480,565){\makebox(0,0)[lb]{\raisebox{0pt}[0pt][0pt]{ $Q$}}}
\put(180,565){\makebox(0,0)[lb]{\raisebox{0pt}[0pt][0pt]{ $P$}}}
\end{picture}}
\end{center}

\paragraph{Composition as Interaction}

\begin{center}
\epsfig{file=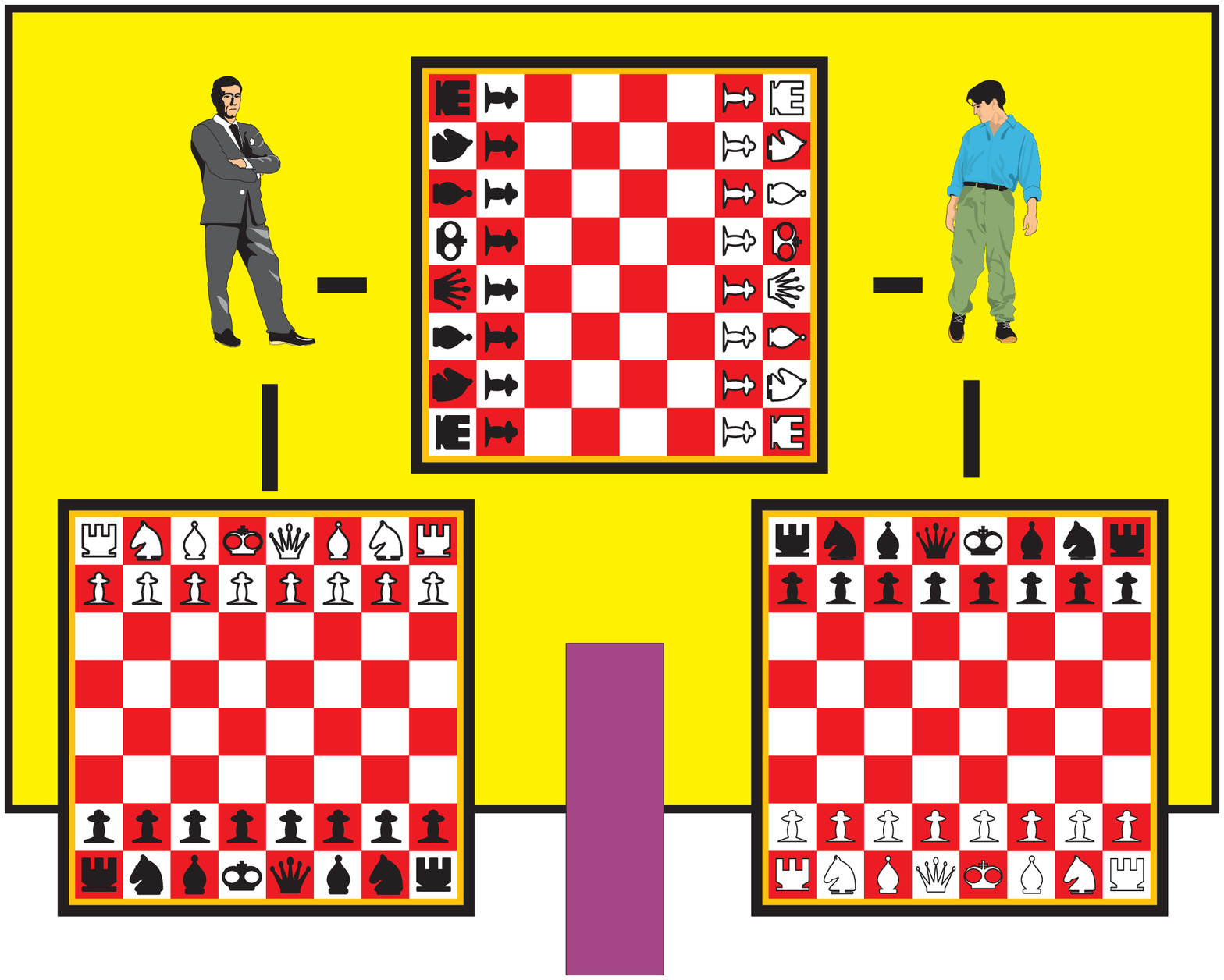,height=2.5in,width=3.5in}\\
The Interaction Game
%\leavevmode\epsfysize=.85\textheight\epsfbox{}
\end{center}
The picture here shows the new system formed by plugging together the two sub-systems. The ``external interface'' to the environment now shows just the left hand board  $A$ as input, and the right hand board $C$ as output. The Cut formula $B$ is hidden from the environment, and becomes the locus of interaction inside the black box of the system. Suppose that the Environment makes some move $m$ in $C$. This is visible only to Nigel, who as a strategy for $B \llto C$ has a response. Suppose this response $m_{1}$ is in $B$. This is a move by Nigel as Player in $B^{\perp}$, hence appears to Gary as a move by  Opponent in $B$. Gary as a strategy for $A \llto B$ has a response $m_{2}$ to this move. If this response is again in $B$, Nigel sees it as a response by the environment to his move, and will have a response again; and so on. We thus have a sequence of moves $m_{1}, \ldots , m_{k}$ in $B$, ping-ponging back and forth between Nigel and Gary. If, eventually, Nigel responds to Gary's last move by playing  in $C$, or Gary responds to Nigel's last move by playing in $A$, then we have the response of the \emph{composed strategy} ${\rm Gary} ; {\rm Nigel}$ to the original move $m$. Indeed, all that is visible to the Environment is that it played $m$, and eventually some response appeared, in $A$ or $C$.

Moreover, if both Nigel and Gary are winning strategies, then so is the composed strategy; and the composed strategy will not get stuck forever in the internal ping-pong in $B$. To see this, suppose for a contradiction that it did in fact get stuck in $B$. Then we would have an infinite play in $B$ following the winning strategy Gary for \emph{Player} in $B$, and the \emph{same} infinite play following the winning strategy Nigel for Player in $B^{\perp}$, hence for \emph{Opponent} in $B$. Hence the same play would count as a win for both Player and Opponent. This yields the desired contradiction.

\subsection{Discussion}
Game Semantics in the sense discussed in this section has had an extensive development over the past decade and a half, with a wealth of applications to the semantics of programming languages, type theories and logics \cite{AJ92b,Abramsky2000,AM97a,AM97b,AM97c,AM99a,AM99b,Hyland2000}. More recently, there has been an algorithmic turn, and some striking applications to verification and program analysis \cite{GM00,Abr2002a,AGMO04,MOW05}.

From the point of view of the general analysis of  Information, we see the following promising lines of development:
\begin{itemize}
\item Game semantics provides a promising arena for exploring the combination of quantitative and qualitative theories of information, as discussed in Section~4, but now in a dynamic setting. In particular, it provides a setting for quantifying information flow between agents. We would like to ask quantitative questions
  about \emph{rate of information flow} through a strategy (representing a
  program, or a proof); how can a system gain \emph{maximum}
  information from its environment while providing \emph{minimal}
  information in return; robustness in the presence of \emph{noise}, etc.
 \item As we saw in our discussion of the copy-cat strategy, there is an intuition of logical principles arising as \emph{conservation laws for information flow}. (And indeed, in the case of Multiplicative Linear Logic, the proofs correspond exactly to ``generalized copy-cat strategies''). Can we develop this intuition into a fully-fledged theory? Can we \emph{characterize} logical principles as those expressing the conservation principles of this information flow dynamics?
\item There is also the hope that the more structured setting of game semantics will usefully constrain the exuberant variety of possibilities offered by process algebra, and allow a sharper exploration of the logical space of possibilities for information dynamics\footnote{It should be said that the exuberant variety of process algebras has been directly motivated by the vast range of real-word informatic processes which we need to model. The whole area of information dynamics is in a dynamic tension between the need on the one hand for descriptive adequacy, and on the other for mathematical structure and tractability \cite{Mil06a,Mil06b}. Process algebra, game semantics, and other approaches are making valuable inroads into this territory. We need to combine the strengths of all these ideas!}. This has already been borne out in part, by the success of game semantics in exploring the space of programming language semantics. It has been possible to give
crisp characterizations of the ``shapes'' of
computations carried out within certain
\emph{programming disciplines}:  including
purely functional programming \cite{Abramsky2000,Hyland2000},
stateful programming \cite{AM97a,AM97b}, general references \cite{AHM98},
programming with non-local jumps and exceptions \cite{Lai97,Lai01}, non-determinism \cite{HaMc99}, probability \cite{DaHa02}, concurrency \cite{GM04,GM06}, names \cite{AGMOS04}, polymorphism \cite{Hug00,AJ05} and more. See \cite{AM99a} for an overview (now rather out of date).

There has also been a parallel line of development of giving \emph{full completeness} results for a range of logics and type theories, characterizing the ``space of proofs'' for a logic in terms of informatic or geometric constraints which pick out those processes which are proofs for that logic \cite{AJ92b,AM99b,Loa,BS98,DHPP,BHS05}. This allows a new look at such issues as the boundaries between classical and constructive logic, or the fine structure of polymorphism and second-order quantification.
\item This also gives some grounds for optimism that  we can capture---in a ``machine-independent'', and moreover ``geometrical'', non-inductive way---what \emph{computational
processes} are, \emph{without} referring back to Turing machines
or any other explicit machine model.
\item In the same spirit as for computability, can we
characterize 
\emph{polynomial-time computation} and other complexity classes in such terms? 
\end{itemize}

\section{Emergent Logic: The Geometry of Information Flow}

Game Semantics carries many vivid intuitions arising from our experiences of game-playing as a human activity. We were able to take advantage of this in the previous section to explain some key ideas without resorting to any explicit formalization. We now turn to a related but somewhat different development of interactive models for logic and computation, known loosely as ``Geometry of Interaction particle-style models''.\footnote{See  \cite{Gi89,Gi90,Gi95,MR91,DR93,DR}, and \cite{Abramsky94a,AJ92b,Abramsky96,Abr97,AbLe,AHS}.}
We will use this setting to carry forward our discussion of dynamic models for information flow, with particular emphasis on the following themes:
\begin{itemize}
\item Firstly, the model or family of models we shall discuss is technically simpler to formalize mathematically than Game Semantics, although also less cloaked in familiar intuitions.
Thus we can introduce some more precision into our discussion without unduly taxing the reader.
\item Secondly, the simple yet expressive nature of these models is itself of conceptual interest. They show how logic and computation can be understood in terms of simple processes of copying information from one ``place'' to another, generalizing what we have already seen of the copy-cat strategy. In fact, we shall see that \emph{mere copying is computationally universal}. Moreover, models of logics and type theories arise from these models; because of the simplicity of the models, we may reasonably speak of \emph{emergent logic}---where, as discussed in the previous section, we may think  of the logical character of certain  principles as arising from the fact that they express conservation laws of information flow.

\item We will also be able to make visible how  geometrical structure unfolds in these models, in  a striking and unexpected fashion. This part of the development can be carried much further than we can describe here; there is a thread of ideas linking logical processes of cut-elimination to diagram algebras, knot theory and topological quantum field theory \cite{Abr07}.

\item We shall also begin to see the beginnings of links between \emph{Logic} and \emph{Physics}.
The processes we shall describe will be \emph{reversible} in a very strong sense.  
This link can in fact be carried much further,  and the same kind of structures we are discussing here can be used to axiomatize Quantum Mechanics, and to give an incisive analysis of quantum entanglement and information flow \cite{Abramsky2002,AbrCoe1,AbrCoe2,Abr07}.
\end{itemize}

\subsection{Background: Combinatory Logic}
It will be convenient to work in the setting of Combinatory Logic \cite{Cur58,HS}, which provides one of the simplest of all the formulations of computability---and moreover one which is purely algebraic. Combinatory Logic is also the basis for realizability constructions, which provide powerful methods for building extensional models of strong impredicative type theories and higher-order logics.

We recall that combinatory logic is the algebraic theory $\CL$ given
by the signature with one binary operation (application) written
as an infix $\_\cdot\_$, and two constants $\SSS$ and $\KK$, subject to 
the equations
\[ \begin{array}{lcl}
\KK \cdot x \cdot y & = & x \\
\SSS \cdot x \cdot y \cdot z & = & x \cdot z \cdot (y \cdot z) 
\end{array} \]
(application associates to the left, so $x \cdot y \cdot z = (x \cdot
y) \cdot z$). Note that we can define $\II \equiv \SSS \cdot \KK \cdot
\KK$, and verify that $\II \cdot x = x$.

The key fact about the combinators is that they are \emph{functionally 
  complete}, i.e. they can simulate the effect of
$\lambda$-abstraction. Specifically, we can define bracket abstraction 
on  combinatory terms built using  a set of variables $X$:
\[ \begin{array}{lcl}
\babs{x}{M} & = & \KK \cdot M \quad (x \not\in \mathsf{FV}(M)) \\
\babs{x}{x} & = & \II  \\
\babs{x}{M \cdot N} & = & \SSS \cdot (\babs{x}{M}) \cdot (\babs{x}{N}) 
\end{array} \]
Moreover (Theorem 2.15 in \cite{HS}):
\[ \CL \vdash (\babs{x}{M}) \cdot N = M[N/x]. \]

\noindent The $\BBB$ combinator can be defined by bracket abstraction from its
defining equation:
\[ \BBB \cdot x \cdot y \cdot z = x \cdot (y \cdot z) .\]
The combinatory \emph{Church numerals} are then defined by
\[ \bar{n} \equiv (\SSS \cdot \BBB )^{n} \cdot (\KK \cdot \II ) \]
where we define
\[ a^n \cdot b = a\cdot (a \cdots (a \cdot b)\cdots ) . \]
A partial function $\phi : \Nat \pfun \Nat$ is \emph{numeralwise
  represented} by a combinatory term $M$ if for all $n \in \Nat$, if
$\phi (n)$ is defined and equal to $m$, then
\[ \CL \vdash M \cdot \bar{n} = \bar{m} \]
and if $\phi (n)$ is undefined, then $M\cdot \bar{n}$ has no normal
form.

\noindent The basic result on computational universality of $\CL$ is then the following (Theorem 4.18 in \cite{HS}):
\begin{theorem}
\label{clcomp}
The  partial functions numeralwise representable in $\CL$ are exactly
the partial recursive functions.
\end{theorem}

\paragraph{Principal Types of Combinators}
The functional behaviour of combinatory terms can be described using \emph{types}. The  type expression $T \rarr U$ denotes the set of terms which, when applied to an argument of type $T$, produce a result of type $U$. By convention, $\rarr$ associates to the right, so we write  $T_{1} \rarr T_{2} \rarr \cdots T_{k }\rarr U$ as short-hand for  $T_{1} \rarr (T_{2} \rarr \cdots (T_{k }\rarr U) \cdots )$.

Now consider the combinator $\KK$. The equation $\KK \cdot x \cdot y  =  x$ tells us that this combinator expects to receive an argument $x$, say of type $\alpha$, then an argument $y$, say of type $\beta$, and then returns a result, namely $x$, of type $\alpha$. Thus its type has the form
\[ \KK :  \alpha \rightarrow (\beta \rightarrow \alpha) . \]
In fact, if we take $\alpha$ and $\beta$ to be type variables, this is the \emph{principal}, \ie the most general, type of this combinator. A similar but more complicated argument establishes that the principal type of the $\SSS$ combinator is 
\[ \SSS  :  (\alpha \rarr \beta \rarr \gamma) \rarr (\alpha \rarr \beta) \rarr (\alpha \rarr \gamma) . \]
These principal types can in fact be computed by  the Hindley-Milner algorithm \cite{Hin} from the defining equations for the combinators. (This algorithm is nowadays routinely used to perform ``type-checking'' for modern programming languages with polymorphic types.)

Curry observed \cite{Cur58} that the principal  types of the combinators correspond to \emph{axiom schemes} for a Hilbert-style proof system for Intuitionistic implicational logic---with the application operation corresponding to \emph{Modus Ponens}. This is the  ``Curry'' part of the Curry-Howard isomorphism.
Thus combinators are to Hilbert-style systems as $\lambda$-calculus is to Natural Deduction.

\paragraph{The Curry Combinators}
Curry's original set of combinators was not the Sch\"onfinkel combinators $\SSS$ and $\KK$, but rather the combinators
$\BB$, $\CC$, $\KK$, and $\WW$:
\[ \begin{array}{lcl}
\BBB \cdot x \cdot y \cdot z & = & x \cdot (y \cdot z) \\
\CCC  \cdot x \cdot y \cdot z & = & x \cdot z \cdot y \\
\WW \cdot x \cdot y & = & x \cdot y \cdot y
\end{array} \]
These combinators are equivalent to the Sch\"onfinkel combinators, in the sense that the two sets are inter-definable  \cite{Bar84,HS}. In particular, $\SSS$ can be defined from $\BBB$,
$\CCC$, $\II$   
and $\WW$. 
They  have the following principal types: 
\[ \begin{array}{lcll}
\II & : & \alpha \rightarrow \alpha & \mbox{Axiom} \\
\BB & : & (\beta \rightarrow \gamma ) \rightarrow (\alpha \rightarrow \beta ) \rightarrow \alpha \rightarrow \gamma & \mbox{Cut} \\
\CC & : & (\alpha \rightarrow \beta \rightarrow \gamma ) \rightarrow \beta \rightarrow \alpha \rightarrow \gamma & \mbox{Exchange} \\
\KK & : & \alpha \rightarrow \beta \rightarrow \alpha & \mbox{Weakening} \\
\WW & : & (\alpha \rightarrow \alpha \rightarrow \beta ) \rightarrow \alpha \rightarrow \beta & \mbox{Contraction} 
\end{array} \]
Thus we see that in logical terms, $\BB$ expresses the transitivity of implication, or the Cut rule; $\CC$ is the Exchange rule; $\WW$ is Contraction; and $\KK$ is Weakening.
Curry's analysis of \emph{substitution} is close to Gentzen's analysis of \emph{proofs}.

\subsection{Linear Combinatory Logic}

We shall now present another system of combinatory logic: \emph{Linear 
  Combinatory Logic} \cite{Abr97,AHS,AbLe}. This can be seen as a finer-grained system into
which standard combinatory logic, as presented in the previous
section, can be interpreted. By exposing some finer structure, Linear
Combinatory Logic offers a more accessible and insightful path towards 
our goal of mapping universal functional computation into a simple model of
computation as copying.

Linear Combinatory Logic can be seen as the combinatory analogue of
Linear Logic \cite{Gi87}; the interpretation of standard Combinatory Logic into
Linear Combinatory Logic corresponds to the interpretation of
Intuitionistic Logic into Linear Logic. Note, however, that the
combinatory systems we are considering are type-free and
``logic-free'' (\textit{i.e.} purely equational).

\begin{definition}\label{LCA}
A {\em Linear Combinatory Algebra} $(A,\ap, \llbang)$ consists of the following 
data:
\begin{itemize}
\item An applicative structure $(A,\ap)$
\item A unary operator $\llbang: A\rightarrow A$
\item Distinguished elements $\BBB$, $\CCC$, $\II$, $\KK$, 
  $\DD$, $\delta$, $\FF$, $\WW$ of $A$
\end{itemize}
satisfying the following identities (we associate $\ap$ to the left and write
$ x  \, \cdot \llbang y$ for $x
\ap (\llbang (y))$, etc.) for all variables $x,y,z$ ranging over $A$):
\[ \begin{array}{clclr}
1. &  \BBB \cdot x \cdot y \cdot z & = &  x \cdot (y \cdot z) &
\mbox{{\rm Composition/Cut}} \\
2. & \CCC \cdot x \cdot y \cdot z & = &  (x \cdot z) \cdot y & \mbox{{\rm Exchange}} \\
3. & \II \cdot x & = &  x & \mbox{{\rm Identity}} \\
4. & \KK \cdot x \, \cdot \llbang y & = &  x & \mbox{{\rm Weakening}} \\
5. & \DD \, \cdot \llbang x & = &  x & \mbox{{\rm Dereliction}} \\
6. & \delta \, \cdot \llbang x & = &  \llbang \llbang x &
  \mbox{{\rm Comultiplication}} \\
7. & \FF \, \cdot \llbang x \, \cdot \llbang y & = &  \llbang (x \cdot
  y) & \mbox{{\rm Monoidal Functoriality}} \\
8. & \WW  \cdot x \, \cdot \llbang y & = &  x \, \cdot \llbang y \, \cdot
  \llbang y & \mbox{{\rm Contraction}} \\
\end{array} \]
\end{definition}
 
\noindent
The notion of LCA corresponds to a Hilbert style
axiomatization of the  $\{!, \linimpl\}$ fragment of
linear logic \cite{Abr97,Avr88,Tro92}. The \emph{principal types} of
the combinators correspond to the axiom schemes which they name. They
can be computed by a Hindley-Milner style algorithm \cite{Hin} from the
above equations:
\[ \begin{array}{llcl}
1. & \BBB & : & (\beta \linimpl \gamma ) \linimpl (\alpha \linimpl \beta ) \linimpl
  \alpha \linimpl \gamma \\
2. & \CCC & : & (\alpha \linimpl \beta \linimpl \gamma ) \linimpl (\beta \linimpl \alpha 
  \linimpl \gamma ) \\
3. & \II & : & \alpha \linimpl \alpha \\
4. & \KK & : & \alpha \linimpl \llbang \beta \linimpl \alpha \\
5. & \DD & : & \llbang \alpha \linimpl \alpha \\
6. & \delta & : & \llbang \alpha \linimpl \llbang \llbang \alpha \\
7. & \FF & : & \llbang (\alpha \linimpl \beta ) \linimpl \llbang \alpha
  \linimpl  \llbang 
\beta \\
8. & \WW & : & (\llbang \alpha \linimpl \llbang \alpha \linimpl \beta ) \linimpl \llbang \alpha
  \linimpl \beta \\
\end{array} \]
Here $\linimpl$ is a \emph{linear function type} (linearity means that 
the argument is used exactly once), and $\llbang \alpha$ allows
arbitrary copying of an object of type $\alpha$.

A {\em Standard Combinatory Algebra}  consists of a pair
$(A,\ap_s)$ where $A$ is a nonempty set and $\ap_s$ is a binary
operation on $A$, together with  distinguished
elements $\BBB_s, \CCC_s, \II_s, \KK_s,$ and $\WW_s$  of $A$,  satisfying
the following identities for all $x,y,z$ ranging over $A$:
\[ \begin{array}{llcl}
1. & \BBB_s\ap_s x\ap_s y \ap_s z & = &  x\ap_s (y\ap_s z) \\
2. & \CCC_s\ap_s x\ap_s y\ap_s z & = &  (x\ap_s z)\ap_s y \\
3. & \II_s \ap_s x & = &  x \\
4. & \KK_s\ap_s x\ap_s  y & = &  x  \\
5. & \WW_s \ap_s x \ap_s y & = &  x\ap_s y\ap_s y \\
\end{array} \]
This is just a combinatory algebra with interpretations of the Curry combinators.
Note that this is equivalent to the more familiar definition of
$\mathbf{SK}$-combinatory algebra as discussed in the previous sub-section.

Let $(A,\ap,\llbang)$ be a linear combinatory algebra. We define a binary
operation $\ap_s$ on $A$ as follows: for $a, b \in A$,
$a \ap_s b \eqdef a \, \ap \llbang b$. 
We define $\DD' \eqdef \mathbf{C \cdot (B \cdot B \cdot I) \cdot (B \cdot D \cdot I)}$. Note that
\[ \DD'  \cdot x  \, \cdot \llbang y = x \cdot y. \]
Now consider
the following elements of $A$.
 
\[ \begin{array}{llcl}
1. & \BBB_s & \eqdef & \mathbf{C\ap (B\ap (B\ap B\ap B)\ap (D'\ap I))\ap (C\ap ((B\ap B)\ap F
)\ap \delta)} \\
2. & \CCC_s & \eqdef & \DD'\ap \CCC \\
3. & \II_s & \eqdef & \DD'\ap \II \\
4. & \KK_s & \eqdef & \DD'\ap \KK \\
5. & \WW_s & \eqdef & \DD'\ap \WW \\
\end{array} \]
 
\begin{theorem}\label{main2}
Let $(A,\ap,\llbang)$ be a linear combinatory algebra. Then $(A, \ap_s)$
with $\ap_s$ and the elements $\BBB_s, \CCC_s, \II_s, \KK_s, \WW_s$ as defined above is a 
standard combinatory algebra.
\end{theorem}

Finally, we mention a special case which will arise in our 
model.
An \emph{Affine Combinatory Algebra} is a Linear Combinatory Algebra
such that the $\KK$ combinator satisfies the stronger equation
\[ \KK \cdot x \cdot y = x. \]
Note that in this case we can \emph{define} the identity combinator: $\II
\eqdef \CCC \cdot \KK \cdot \KK$.

\subsection{Universal Computation by Copy-Cats}

Our aim is to describe an interactive model for logic and computation, which can be understood in two complementary ways:
\begin{itemize}
\item A model built from simple dynamical processes of copying information from one place to another.
\item A model built from simple geometrical constructions, in which computation is interpreted as geometric simplification---tracing paths through tangles, and yanking them straight.
\end{itemize}
We begin with the dynamical interpretation. Here we think of an informatic \emph{token} or \emph{particle} traversing a path through logical (discrete) ``space'' and ``time''. For this purpose, we assume a set $\Pos$ of \emph{positions} or \emph{places} in ``logical space''. For the purposes of obtaining a type-free universal model of computation, it is important that $\Pos$ is (countably) infinite. (So we could just take it to be the set $\Nat$ of natural numbers). The only significant property of the instantaneous state of the particle is its current position $p \in \Pos$.

The \emph{processes} we shall consider will be very simple, ``history-free'' or ``time-independent'' reversible dynamics, which we represent as \emph{partial injective functions}
\[ f : \Pos \rightharpoonup \Pos . \]
Such a process maps a particle in position $p$ at any time $t$ to the position $f(p)$ at time $t+1$; or may be undefined. In fact, we will have no need to make time explicit, since discrete time will be modelled adequately by \emph{function composition}.\footnote{The non-trivial dynamics we shall actually consider, which will arise when we model function application, will in fact come from the interaction between \emph{two} very simple functions --- the ``ping-ponging'' back and forth between them, in the terms of our informal discussion of interaction in Section~5. Note that \emph{all} elements of our combinatory algebra, whether they appear as ``functions'' or ``arguments'' in the context of a given application,   will be represented by functions on positions, corresponding to processes or strategies.}Thus the path traced by the particle starting from position $p_{0}$ under the dynamics $f$ will be
\[ p_{0}, p_{1}, p_{2}, \ldots, p_{n}, \ldots \]
where $p_{i+1} = f(p_{i})$. This dynamics is clearly reversible. Since $f$ is a partial injective map, its inverse $f^{-1}$ (\ie the relational converse of $f$) is also a partial injective function on $\Pos$, and $p_{i} = f^{-1}(p_{i+1})$, so we can trace the reverse path using the inverse dynamics.

In fact, it will be possible to restrict ourselves to an even simpler class of dynamics: namely the \emph{fixed-point free partial involutions}, \ie those partial  injective functions $f : \Pos \pfn \Pos$ satisfying 
\[ f = f^{-1}, \qquad \qquad f \; \cap \; 1_{\Pos} = \varnothing . \]
Thus such a map satisfies:
\[ f(x) = y \;\; \Longleftrightarrow \;\; x = f(y), \qquad \qquad f(x) \neq x . \]

A partial involution on a set $X$ is equivalently described as a partial partition of $X$ into 2-element subsets:
\[
X \supseteq \bigcup E , \qquad \qquad \text{where} \; E = \{ \{ x, y \} \mid f(x) = y \} . 
\]
This defines an undirected graph $G_f = (X, E)$. Clearly each vertex in this graph has at most one incident edge. Conversely, every  graph $G = (X, E)$ with this property determines a unique partial involution $f $ on $X$, with $G_f = G$.

Partial involutions will be our model at this basic level of ``copy-cat processes''; they simply copy information back and forth between pairs of ``locations''.
It is somewhat remarkable that such simple maps can form  a universal computational model.

\subsubsection{Function Application as Interaction}

Our next and key step is to  model  \emph{functional application} by \emph{interaction} of these simple dynamical processes. This will in fact be a bare-bones version of the game-theoretic model of composition as interaction which we gave in the previous section. We shall view a  ``functional process'' which can be applied to other processes as a two-input two-output function
\begin{center}
\psset{unit=1in,cornersize=absolute,dimen=middle}%
\begin{pspicture}(0,-0.75)(1.125,0.75)%
% dpic version 08.Jul.05 for PSTricks 0.93a
\psframe[fillstyle=solid,fillcolor=yellow,linecolor=blue](0,-0.375)(1.125,0.375)
\rput(0.5625,0){$f$}
\psline[arrowsize=0.05in 0,arrowlength=2,arrowinset=0]{<-}(0.28125,0.375)(0.28125,0.75)
\psline[arrowsize=0.05in 0,arrowlength=2,arrowinset=0]{<-}(0.84375,0.375)(0.84375,0.75)
\psline[arrowsize=0.05in 0,arrowlength=2,arrowinset=0]{->}(0.28125,-0.375)(0.28125,-0.75)
\psline[arrowsize=0.05in 0,arrowlength=2,arrowinset=0]{->}(0.84375,-0.375)(0.84375,-0.75)
\end{pspicture}%

\end{center}
The interpretation of these two pairs of input-output lines is that the first will be used to connect the functional process to its argument, and the second to connect it to its external environment or context---which will interact with the function to consume its output.
Formally, this is a function
\[ f : \Pos + \Pos \; \longrightarrow \Pos + \Pos . \]
Note that we have the used the disjoint union (two copies of $\Pos$) rather than the cartesian product $\Pos \times \Pos$ (infinitely many copies of $\Pos$). This is because a particle coming in as input must \emph{either} be on the first input line, \emph{or} (in the exclusive sense) on the second input line; and similarly for the outputs.

However, since we want to make a type-free universal model of computation, we must reduce \emph{all} our processes to one-input one-output functions. This is where our assumption that $\Pos$ is infinite  becomes important. It allows us to define a 
{\em splitting function} $\Split$:
\[ \Split : \Pos + \Pos \rTo^{\cong} \Pos .\] 
We can think of this as splitting logical space into two disjoint ``address spaces''.
This  allows us to
transform any one-input/one-output function into a
two-input/two-output function, or conversely, by conjugation. Thus if $f : \Pos \rarr \Pos$ is any process, we can view it as a two-input, two-output functional process, namely 
\[ \Split^{-1} \circ f \circ \Split : \Pos + \Pos \; \longrightarrow \Pos + \Pos . \]

\subsubsection{Geometrical Representation of Application}
Suppose we wish to apply $f$, \textit{qua} functional process, to $g$, where both $f$ and $g$ are partial involutions on $\Pos$. The application operation $f \bullet g$ is indicated pictorially as follows.
\begin{center}
\begin{picture}(160,120)
\put(50,80){\vector(0,-1){30}}
\put(50,30){\line(0,1){20}}

\put(50,50){\line(3,-2){30}}

\put(80,80){\vector(0,-1){30}}
\put(80,30){\line(0,1){20}}

\put(80,50){\line(-3,-2){30}}

\put(40,30){\framebox(50,20){}}

%\put(40,-20){\makebox(50,20){$(ii)$}}

\put(10,30){\makebox(15,10){$f_{11}$}}
\put(25,35){\line(1,0){23}}

\put(10,40){\makebox(15,10){$f_{12}$}}
\put(25,45){\line(1,0){29}}

\put(105,30){\makebox(15,10){$f_{22}$}}
\put(105,35){\line(-1,0){23}}

\put(105,40){\makebox(15,10){$f_{21}$}}
\put(105,45){\line(-1,0){29}}

\put(50,30){\vector(0,-1){30}}
\put(80,30){\vector(0,-1){30}}

% fig. i

\put(-45,80){\vector(0,-1){30}}
\put(-60,30){\framebox(30,20){$f$}}
\put(-45,30){\vector(0,-1){30}}

%\put(-60,-20){\makebox(30,20){$(i)$}}

%fig. iii

\put(175,115){\vector(0,-1){30}}
\put(205,115){\vector(0,-1){30}}

\put(208,105){\makebox(10,10){$p$}}

\put(165,65){\framebox(50,20){$\Split^{-1}\circ f \circ \Split$}}
\put(175,65){\vector(0,-1){30}}
\put(205,65){\vector(0,-1){30}}

\put(216,35){\makebox(20,10){$f\bullet g (p)$}}

\put(165,15){\framebox(20,20){$g$}}
\put(175,15){\line(0,-1){15}}
\put(175,0){\line(-1,0){20}}

\put(155,0){\line(0,1){115}}
\put(155,115){\line(1,0){20}}

%\put(165,-20){\makebox(50,20){$(iii)$}}

\end{picture}

\end{center}
As already explained, we conjugate $f$ by $\Split$ to turn it into something with the right shape to be a functional process. Then we connect it to its argument, $g$, by a feedback loop using the first input and output lines of $f' = \Split^{-1} \circ f \circ \Split$. The residual behaviour by which the process resulting from the application communicates with its environment uses the second input and output lines. The full geometric significance of how this notion of application works will become apparent when we discuss the interpretation of the combinators in this setting. But we can give the dynamical interpretation of application immediately. Suppose the process $f \bullet g$ receives a token $p$ on its input. The function $f'$ may immediately dispatch this to its second output line as $p'$---in which case, that will be the response of $f\bullet g$. This would correspond to the behaviour of a constant function, which knows its output without consulting its input. Otherwise, $f'$ may dispatch $p$ to its \emph{first} output line, as $p_{1}$. This is then fed as input to $g$. Thus this corresponds to the function represented by $f'$  interrogating its argument. If $g(p_{1}) = p_{2}$, then this is fed back around the loop as input to $f'$ (now on its first input line). We may continue in this fashion, ping-ponging between $f'$ (on its first input/output lines) and $g$ around the feedback loop. Eventually, $f'$ may have seen enough, and decide to despatch the token on its second output line, as $p'$. We then say that $f\bullet g(p) = p'$. In other words, to the external environment, the whole interaction between $f'$ and $g$ has been hidden inside the black box of the application $f \bullet g$; it only sees the final response $p'$ to the initial entry of the token at $p$.

All of this should seem very familiar. It follows exactly the same general lines as the game-semantical interpretation of composition which we presented in the previous section. We note the following points of difference:
\begin{itemize}
\item The notion of composition we discussed in the previous section was fully symmetrical between the two agents involved, reflecting the classical nature of the underlying logic. Here, we are discussing \emph{functional computation}, and our description of application reflects the asymmetry between function and argument.
\item Since we are dealing with a type-free universal model of computation, we must allow some partiality in our model. The token may get trapped in the feedback loop for ever, for example, so the involutions giving the dynamics must be partial in general. This is unavoidable, for well-known metamathematical reasons.
\item We are also considering a very restricted, simple notion of dynamics here. Certainly in the game semantics context, we would not want in general to limit ourselves to such a restricted class of strategies.
\end{itemize}

\subsubsection{Algebraic Description of Application}
We now give a formal definition of the application operation. 
Firstly, consider the map $f' = \Split^{-1} \circ f \circ \Split: \Pos + \Pos \longrightarrow \Pos + \Pos$. Each input lies in \emph{either} the first component of the disjoint union, \emph{or}  (exclusive or) the second, and similarly for the corresponding output. This leads to a decomposition of $f'$ into four \emph{disjoint partial maps} $f_{ij}$, $i, j \in \{ 1, 2\}$, where $f_{ij}$ maps the $i$'th input summand to the $j$'th output summand.
Note that $f'$ can be recovered as the  union of these four maps. 
Since $f'$ is a partial involution,  these maps will also be partial involutions. The decomposition is indicated pictorially  in the preceding diagram. Now we can define
\[ 
 f \bullet g = f_{22} \cup f_{21};
 g; (f_{11}; g)^{\ast} ; f_{12} \, ,
\]
where we  use relational algebra (union $R \cup S$, relational composition $R ; S$ and reflexive transitive closure $R^{\ast}$) to write down formally exactly the information flow we described in our informal explanation of application above. It is a nice exercise to show that partial involutions are closed under application; that is, that $f \bullet g$  is again a partial involution.

\subsection{Combinators as Copy-Cats}
At this point, we have defined our applicative structure $(A, \bullet)$, where $A$ is the set of partial involutions. We must now show that we can  define combinators as partial involutions such that this structure will indeed form a Linear Combinatory Algebra.  From now on, we shall mainly resort to drawing pictures, rather than writing algebraic expressions.

\subsubsection{The Identity Combinator}
Our first example is the simplest, and yet already shows the essence of the matter. The identity combinator $\II$ is represented by the \emph{twist map}, which copies any token on its first input line to the second output line, and vice versa. This is depicted  as follows.
\begin{center}
\begin{picture}(50,120)

\put(-60,45){\makebox{${\textbf I}$}}
\put(5,95){\makebox(10,10){$x$}}

\put(10,90){\vector(0,-1){30}}

\put(10,60){\line(2,-3){20}}

\put(30,90){\vector(0,-1){30}}

\put(30,60){\line(-2,-3){20}}

\put(0,30){\framebox(40,30){}}

\put(10,30){\vector(0,-1){30}}
\put(30,30){\vector(0,-1){30}}

\end{picture}
\end{center}
 What is surprising, and striking, is the geometric picture of \emph{why} this works: that is, why the equation $\II \bullet x = x$ holds:
\begin{center}
\begin{picture}(320,135)
\put(65,115){\vector(0,-1){30}}
\put(95,115){\vector(0,-1){30}}

\put(55,65){\framebox(50,20){}}
\put(65,65){\vector(0,-1){30}}
\put(95,65){\vector(0,-1){30}}

\put(65,85){\line(3,-2){30}}  
\put(95,85){\line(-3,-2){30}}

\put(55,15){\framebox(20,20){$x$}}
\put(65,15){\line(0,-1){15}}
\put(65,0){\line(-1,0){20}}

\put(45,0){\line(0,1){115}}
\put(45,115){\line(1,0){20}}

%\end{picture}

\put(160,30){\mbox{$=$}}

%\put(145,-30){\mbox{$\II x = x$}}

%\begin{picture}(160,100)
\put(240,80){\vector(0,-1){30}}
\put(225,30){\framebox(30,20){$x$}}
\put(240,30){\vector(0,-1){30}}
\end{picture}\\
$\II \bullet x = x$
\end{center} 
We see that geometrically, this is a matter of \emph{yanking the string straight}; while dynamically, we picture the token flowing once around the feedback loop, and exiting exactly according to $x$.

Once again, we can recognize this combinator as a new description of an old friend from the previous section. \emph{This is exactly the copy-cat strategy!} Reduced to its essence, it simply copies ``tokens'' or ``moves'' from one place to another, and \textit{vice versa}; the logical requirement is that one of these places should be positive (or output); while the other should be negative (or input).

\subsubsection{The Composition Combinator}
We now consider the composition combinator $\BB$. We interpret it as a \emph{combination of copy-cats}. That is, it plays copy-cat between three pairs  of input and output lines. (Thus, in particular, it is a partial involution).
\begin{center}
\begin{picture}(130,120)

\put(-60,45){\makebox{${\textbf B}$}}

\put(0,30){\framebox(120,30){}}

%inside connections:

\put(10,30){\line(2,3){20}}
\put(30,30){\line(-2,3){20}}

\put(50,30){\line(2,3){20}}
\put(70,30){\line(-2,3){20}}

\put(90,30){\line(2,3){20}}
\put(110,30){\line(-2,3){20}}

% output and input wires of B:

\put(10,30){\vector(0,-1){30}}
\put(30,30){\vector(0,-1){30}}
\put(50,30){\vector(0,-1){30}}
\put(70,30){\vector(0,-1){30}}
\put(90,30){\vector(0,-1){30}}
\put(110,30){\vector(0,-1){30}}

\put(5,95){\makebox(10,10){$z$}}

\put(30,94){\makebox(20,5){$\overbrace{\ \ \ \ \ \ \  }^y$}}

\put(70,94){\makebox(20,5){$\overbrace{\ \ \ \ \ \ \  }^x$}}

\put(10,90){\vector(0,-1){30}}

\put(30,90){\vector(0,-1){30}}
\put(50,90){\vector(0,-1){30}}

\put(70,90){\vector(0,-1){30}}
\put(90,90){\vector(0,-1){30}}
\put(110,90){\vector(0,-1){30}}

\end{picture}
\end{center}
Note that we can regard this combinator as having six inputs and six outputs, as shown in the diagram, simply by iterating the trick of conjugating it by the splitting map $\Split$. Our reason for giving it this many inputs and outputs is based on the \emph{functionality} of $\BB$, \ie its principal type. It expects to get three arguments, the first two of which will themselves be applied to arguments, and hence should each have two inputs and two outputs.

Once again, the real insight as to how this combinator works will come from the geometry, or equivalently the particle dynamics.
We let the picture speak for itself.

\begin{center}
\begin{picture}(310,240)

\put(50,120){\framebox(120,30){}}

\put(49,70){\framebox(22,20){$z$}}
\put(79,70){\framebox(22,20){$y$}}
\put(119,70){\framebox(22,20){$x$}}

%inside connections:

\put(60,120){\line(2,3){20}}
\put(80,120){\line(-2,3){20}}

\put(100,120){\line(2,3){20}}
\put(120,120){\line(-2,3){20}}

\put(140,120){\line(2,3){20}}
\put(160,120){\line(-2,3){20}}

% output and input wires of B:

\put(60,120){\vector(0,-1){30}}
\put(80,120){\vector(0,-1){30}}
\put(100,120){\vector(0,-1){30}}
\put(120,120){\vector(0,-1){30}}
\put(140,120){\vector(0,-1){30}}
\put(160,120){\vector(0,-1){30}}

\put(60,180){\vector(0,-1){30}}
\put(80,190){\vector(0,-1){40}}
\put(100,200){\vector(0,-1){50}}
\put(120,210){\vector(0,-1){60}}
\put(140,220){\vector(0,-1){70}}
\put(160,180){\vector(0,-1){30}}

%parallel input lines:

\put(40,180){\line(1,0){20}}
\put(30,190){\line(1,0){50}}
\put(20,200){\line(1,0){80}}
\put(10,210){\line(1,0){110}}
\put(0,220){\line(1,0){140}}

%parallel output lines:

\put(40,40){\line(1,0){20}}
\put(30,30){\line(1,0){50}}
\put(20,20){\line(1,0){80}}
\put(10,10){\line(1,0){110}}
\put(0,0){\line(1,0){140}}

%vertical lines:

\put(40,40){\line(0,1){140}}
\put(30,30){\line(0,1){160}}

\put(20,20){\line(0,1){180}}
\put(10,10){\line(0,1){200}}
\put(0,0){\line(0,1){220}}

%output xyz

\put(60,70){\line(0,-1){30}}
\put(80,70){\line(0,-1){40}}
\put(100,70){\line(0,-1){50}}
\put(120,70){\line(0,-1){60}}
\put(140,70){\line(0,-1){70}}

%\end{picture}
%\end{center}

%

%
%\begin{center}
%\begin{picture}(80,180)

\put(249,25){\framebox(22,20){$z$}}
\put(259,75){\framebox(22,20){$y$}}
\put(279,125){\framebox(22,20){$x$}}

% wires of x

\put(280,175){\vector(0,-1){30}}
\put(300,175){\vector(0,-1){30}}

\put(280,125){\vector(0,-1){30}}
\put(300,125){\vector(0,-1){30}}

% wires of y

\put(260,110){\vector(0,-1){15}}

\put(260,75){\vector(0,-1){30}}
\put(280,75){\line(0,-1){75}}

% wire of z

\put(260,25){\line(0,-1){15}}

%vertical

\put(240,10){\line(0,1){100}}
\put(230,0){\line(0,1){175}}

%parallel

\put(230,0){\line(1,0){50}}
\put(240,10){\line(1,0){20}}
\put(240,110){\line(1,0){20}}
\put(230,175){\line(1,0){50}}

%\put(220,80){\mbox{$=$}}

%\put(160,0){\mbox{$\BB x y z = x (yz)$}}

\end{picture}\\
\vspace{.2in}
$\BB \cdot x \cdot y \cdot z = x \cdot (y\cdot z)$
\end{center}

\subsubsection{Other Affine Combinators}

The remaining Linear Combinators can be described in similar style.
We simply show the definition for $\CC$.
\begin{center}
\begin{picture}(130,120)

\put(-60,45){\makebox{${\textbf C}$}}

\put(0,30){\framebox(120,30){}}

%inside connections:

\put(10,30){\line(2,1){60}}
\put(30,30){\line(2,3){20}}

\put(50,30){\line(-2,3){20}}
\put(70,30){\line(-2,1){60}}

\put(90,30){\line(2,3){20}}
\put(110,30){\line(-2,3){20}}

% output and input wires of C:

\put(10,30){\vector(0,-1){30}}
\put(30,30){\vector(0,-1){30}}
\put(50,30){\vector(0,-1){30}}
\put(70,30){\vector(0,-1){30}}
\put(90,30){\vector(0,-1){30}}
\put(110,30){\vector(0,-1){30}}

\put(5,95){\mbox{$z$}}

\put(25,95){\mbox{$y$}}

%\put(55,94){\makebox(30,10){$\overbrace{\ \ \ \ \ \ \  }^x$}}
\put(50,94){\makebox(40,5){$\overbrace{\ \ \ \ \ \ \ \ \ \ \ \ \ \  }^x$}}

\put(10,90){\vector(0,-1){30}}

\put(30,90){\vector(0,-1){30}}
\put(50,90){\vector(0,-1){30}}

\put(70,90){\vector(0,-1){30}}
\put(90,90){\vector(0,-1){30}}
\put(110,90){\vector(0,-1){30}}

\end{picture}\\
\vspace{.2in}
$\CC \cdot x \cdot y \cdot z = x \cdot z \cdot y$
\end{center}
We note that geometrically, this is our first example of a \emph{non-planar} combinator.\footnote{Our diagrammatic conventions obscure this point, since all our diagrams involve over-crossing lines. For an explicit discussion of planarity and an alternative diagrammatics, see \cite{Abr07}.}  This gives a hint of the geometrical possibilities lurking just below the surface. We shall not pursue this fascinating theme here for lack of space, but see \cite{Abr07}.

In fact, the algebra is naturally \emph{affine}. We can define a $\KK$ combinator:
\begin{center}
\begin{picture}(70,120)

\put(-60,45){\makebox{${\textbf K}$}}

\put(0,30){\framebox(60,30){}}

%inside connections:

\put(30,30){\line(2,3){20}}
\put(50,30){\line(-2,3){20}}

% output and input wires of K:

\put(10,30){\vector(0,-1){30}}
\put(30,30){\vector(0,-1){30}}
\put(50,30){\vector(0,-1){30}}

\put(10,90){\vector(0,-1){30}}
\put(30,90){\vector(0,-1){30}}
\put(50,90){\vector(0,-1){30}}

\put(5,95){\makebox(10,10){$y$}}
\put(25,95){\makebox(10,10){$x$}}

\end{picture}\\
\vspace{.2in}
$\KK \cdot x \cdot y = x$
\end{center}
However, note that another topological property is violated here; the first input and output lines are \emph{disconnected} from the information flow. (Recall our discussion of the second variation of the Chess copy-cat scenario).

\subsubsection{Duplication}
We shall conclude our discussion of the algebra by sketching how explicit duplication of arguments is handled. This is needed for full expressive power.

We define another auxiliary function
\[ \pair : \Nat \times \Pos \rTo^{\cong} \Pos \] 
which splits logical space into countably many disjoint copies. Again, this requires the assumption that $\Pos$ is infinite. Using this, we can define an
operation $\bang  f$ which is intended to produce {\em infinitely many
copies of $f$}. These are obtained by simply tagging each  copy with a natural
number, i.e. we define:
\[ \bang f = \pair \circ (1_{\Nat} \times f) \circ \pair^{-1} .\]
We can then define $\WW$ satisfying
\[ \WW \cdot x \cdot \bang y = x \cdot \bang y \cdot \bang y . \]

\paragraph{The $\WW$ combinator}
\begin{center}
\begin{picture}(170,120)

\put(-60,45){\makebox{$\WW'$}}

\put(0,30){\framebox(160,30){}}

%inside connections:

\put(30,30){\line(2,1){60}}
\put(50,30){\line(2,1){60}}

\put(90,30){\line(-2,1){60}}
\put(110,30){\line(-2,1){60}}

\put(130,30){\line(2,3){20}}
\put(150,30){\line(-2,3){20}}

% output and input wires of W:

\put(30,30){\vector(0,-1){30}}
\put(50,30){\vector(0,-1){30}}

\put(90,30){\vector(0,-1){30}}
\put(110,30){\vector(0,-1){30}}

\put(130,30){\vector(0,-1){30}}
\put(150,30){\vector(0,-1){30}}

\put(8,98){\makebox(70,5){$\overbrace{\ \ \ \ \ \ \ \ \  \ \ \ \ \ \ \ \ \ \ \ \ \ \ \  }^{\bang y}$}}

\put(90,94){\makebox(40,5){$\overbrace{\ \ \ \ \ \ \ \ \ \ \ \ \ \  }^x$}}

\put(30,90){\vector(0,-1){30}}
\put(50,90){\vector(0,-1){30}}

\put(90,90){\vector(0,-1){30}}
\put(110,90){\vector(0,-1){30}}

\put(130,90){\vector(0,-1){30}}
\put(150,90){\vector(0,-1){30}}

\put(10,80){\makebox(10,10){$(\mathbf{l}.i,n)$}}
\put(62,80){\makebox(10,10){$(\mathbf{r}.j,m)$}}

\put(80,-10){\mbox{$(i, n)$}}
\put(110,-10){\mbox{$(j, m)$}}

\put(10,70){\makebox(20,10){$\ldots$}}
\put(50,70){\makebox(20,10){$\ldots$}}

\put(10,10){\makebox(20,10){$\ldots$}}
\put(50,10){\makebox(20,10){$\ldots$}}

\end{picture}
\end{center}
\vspace{.2in}
This combinator can be understood as effecting a
``translation between dialects'':
\begin{itemize}
\item $x$ sees two arguments, each in many copies.
\item $\bang y$ provides one argument, in as many copies as needed.
\end{itemize}
The combinator in effect decomposes into infinitely many copy-cat strategies, using a suitable splitting function to  split the ``address space'' of the countably many copies of $\bang y$ into two infinite, disjoint parts,  and copying between each of these and the corresponding argument position of $x$.

\subsection{Putting the Pieces Together}
We can round out the descriptions of the combinators as partial involutions to obtain a Linear Combinatory Algebra. By Theorem~\ref{main2}, this yields a standard Combinatory Algebra, and hence by Theorem~\ref{clcomp} a universal model of computation. Moreover, realizability constructions over this Combinatory Algebra provide models for higher-order logics and type  theories. Thus we have fulfilled our programme for this Section, of exhibiting the power of copying, leading to emergent models of logic and computation.

\subsection{Discussion}
Our gentle description of the partial involutions model in this section has merely indicated some first steps in this topic. We list some further directions:
\begin{itemize}
\item There is a general axiomatic formulation of this construction in terms of \emph{traced monoidal categories} \cite{JSV}, with instances for deterministic, non-deterministic, probabilistic and quantum interaction \cite{Abramsky96,AHS}.
\item The connections with reversible computation have been mentioned; this topic is carried further in \cite{Abrrev}.
\item These models have some striking applications to the analysis of proofs, and of definability in various type theories, via \emph{Full Completeness theorems} for models arising by realizability constructions over the basic Geometry of Interaction models \cite{AbLe}.
\item Current work is showing that the suggestive connections with geometry can be carried much further. In particular, there are connections with diagram algebras such as the Temperley-Lieb algebra, and hence with the Jones polynomial and ensuing developments \cite{Abr07}.
\item Finally, as already mentioned, there are strong connections with Quantum Information and Computation, which deserve a proper account of their own. Some references are \cite{Abramsky2002,AbrCoe1,AbrCoe2}.
\end{itemize}

%\section{Quantum Entanglement and Information Flow}

\section{Concluding Remarks}

The underlying project we have tried to illuminate in this article, via some partial exemplifications, has been that of developing a free-standing, syntax-independent Information Dynamics, worthy of the name.

In our view, this project is significant not just for Computer Science, but for Applied Logic, and for the theory and philosophy of information in general.

\subsection{Combining Static and Dynamic}
We have already emphasized the importance of combining qualitative and quantitative aspects of information, in the context of both static and dynamic theories. We conclude by making the point that it can be fruitful to combine static and dynamic aspects as well.\footnote{This point was emphasized by Johan van Benthem (personal communication).}

We can set this in a wider context.
One can distinguish two views on how Logic relates to Structure:
\begin{enumerate}
\item \textbf{The Descriptive View.} Logic is used to \emph{talk about} structure.
This is the view taken in Model Theory, and in most of the uses of Logic (Temporal logics, MSO etc.) in Verification in Computer Science.
It is by far the more prevalent and widely-understood view.
\item \textbf{The Intrinsic View.} Logic is taken to \emph{embody} structure. This is, implicitly or explicitly, the view taken in the Curry-Howard isomorphism, and more generally in Structural Proof Theory, and in (much of) Categorical Logic. In the Curry-Howard isomorphism, one is not using logic to \emph{talk about} functional programming; rather, logic (in this aspect)  \emph{is} functional programming.
\end{enumerate}

The descriptive view is well exemplified by Dynamic Logic and other modal logics. Indeed, one can use modal logics to \emph{talk about} games and strategies, while on the other hand these can be used as a manifestation of the intrinsic view, modelling proofs as certain interactive processes. In some sense the intrinsic view is \emph{global}, giving the structure of a type theory or semantic category; while the descriptive view is \emph{local}, exploring the structure of particular games (objects) or strategies (morphisms). There is no reason why these two views cannot be brought fruitfully together, e.g.~using a suitable modal logic to verify the logical soundness properties of strategies such as the copy-cat strategies we have discussed.

This further attempt to draw some of the strands we have examined in this article together is one of many promising directions for future work.

\subsection{The Fundamental Challenge}
The most fundamental challenge faced by the project of an Informatic Dynamics is in our view this: how to expose what is really robust and intrinsic structure, a bedrock on which we can build, as opposed to what is more or less arbitrarily chosen.\footnote{There is a name  in Computer Science for the syndrome of a  profusion of choices, none canonical: the ``next 700 syndrome''. It comes from Peter Landin's classic paper (from 1966!), ``The Next 700 Programming Languages'' \cite{Lan}. For further discussion of this syndrome, and where we might find inspiration in addressing it, see \cite{Abr06}.}
This problem is all the more acute, given the ever-increasing range of concrete informatic phenomena which we are continually being challenged to model by the  rapidly-moving world of Information Technology.

Without under-estimating these difficulties, we find numerous, if ``local'', grounds for optimism in the theories we have surveyed, in the insights and structures which they have brought to light.
We venture to believe that real and exciting progress will continue to be made, and that a fundamental and widely applicable scientific theory of Information, incorporating qualitative and structural as well as quantitative features,  is in the making.

%}
\end{document}